\documentclass{aa}  
\usepackage{graphicx}

\usepackage{txfonts}
\usepackage{natbib}
\usepackage{rotating}
\bibpunct{(}{)}{;}{a}{}{,} 

\usepackage{hyperref}
\newcommand{\DTM}[0]{$\mathcal{DT}\hspace{-0.03in}\mathcal{M}$ }
\begin{document}

   \title{Evolution of the dust-to-metals ratio in high-redshift galaxies probed by GRB-DLAs\thanks{Based on observations collected at the European Southern Observatory, Paranal, Chile, Program IDs: 088.A-0051(B), 089.A-0067(B), 091.C-0934, 094.A-0134(A)}}


   \author{P. Wiseman
        \inst{1}
          \and
         P. Schady
        \inst{1}
        \and
          J. Bolmer
         \inst{1}
          \and
        T. Kr{\"{u}}hler
         \inst{1}
        \and
        R. M. Yates
        \inst{1}
        \and
        J. Greiner
        \inst{1}
        \and
        J. P. U. Fynbo
        \inst{2}
         }

   \institute{Max-Planck-Institute f\"ur Extraterrestrische Physik (MPE), Giessenbachstrasse 1, 85748 Garching, Germany\\
        \email{wiseman@mpe.mpg.de}
        \and
          Dark Cosmology Centre, Niels Bohr Institute, University of Copenhagen, Juliane Maries Vej 30, 2100 Copenhagen, Denmark
        }
   \date{Received 1  July 2016/Accepted October 26 2016}

 
  \abstract
   {Several issues regarding the nature of dust at high redshift remain unresolved: its composition, its production and growth mechanisms, and its effect on background sources. }
   {We provide a more accurate relation between dust depletion levels and dust-to-metals ratio (DTM), and to use the DTM to investigate the origin and evolution of dust in the high-redshift Universe via Gamma-ray burst damped Lyman-alpha absorbers (GRB-DLAs).}
   {We use absorption-line measured metal column densities for a total of 19 GRB-DLAs, including five new GRB afterglow spectra from VLT/X-shooter. We use the latest linear models to calculate the dust depletion strength factor in each DLA. Using these values we calculate total dust and metal column densities to determine a DTM. We explore the evolution of DTM with metallicity, and compare it to previous trends in DTM measured with different methods. }
   {We find significant dust depletion in 16 of our 19 GRB-DLAs, yet 18 of the 19 have a DTM significantly lower than the Milky Way. We find that DTM is positively correlated with metallicity, which supports a dominant ISM grain-growth mode of dust formation. We find a substantial discrepancy between the dust content measured from depletion and that derived from the total $V$-band extinction, $A_V$, measured by fitting the afterglow SED. We advise against using a measurement from one method to estimate that from the other until the discrepancy can be resolved.}
   {}


   \maketitle
%

\section{Introduction}

The abundances and compositions of the dust and metals in the interstellar medium (ISM) can reveal important information about  local environmental conditions. Despite the wealth of information on our doorstep regarding the ISM of the Milky Way (MW) and Local Group galaxies, it is also necessary to investigate the ISM in the distant Universe in order to trace its properties in very different environments, as well as its evolution over cosmic history. 

One of the key constituents of the ISM is dust. Dust is produced in a range of environments, from the stellar sources of outer envelopes of post-aysmptotic giant branch (AGB) stars and the expanding and cooling ejecta of supernova to grain growth and accretion in the ISM. It reveals itself  via emission in the far-infrared and sub-mm wavelength range and  through absorption and scattering of visible and ultraviolet (UV) light from background sources, and its effect must be corrected for when studying sources that shine through it. For example, everything outside the Galaxy must be observed through the dust of the MW, which has a complex topography \citep{Schlafly2011}. It is estimated that up to 30\% of all light in the Universe has been reprocessed by dust grains \citep{Bernstein2002}. Dust is also necessary for, and traces, star formation across the Universe \citep{Sanders1996,Genzel1998,Peeters2004,McKee2007}. Conversely, star formation also destroys dust at differing rates (\citealt{Draine1979b,Draine1979a,McKee1989,Jones1996,Dwek1998,Bianchi2005,Yamasawa2011}). 
Along with the ISM, dust is present in substantial quantities alongside gas and metals in the circum-galactic medium (CGM; \citealt{Bouche2007,Peeples2014,Peek2015}.)
It is therefore of fundamental importance to the theory of star formation and thus galaxy evolution to understand the nature of all dust processes, such as formation, composition, evolution, and destruction, as well as its observational characteristics, both in the local and distant Universe. Since dust is intimately connected to the conditions of the ISM and the properties of gas \citep{Draine2003}, the dust-to-gas ratio (DTG; \citealt{Bohlin1978}) is a good indicator of the dust content of a galaxy or gas cloud. The dust-to-metals ratio (DTM; \citealt{Predehl1995,Guever2009,Watson2011}), which is the DTG corrected for the metallicity of the gas, thus describing the fraction of the total metals that are in the solid dust phase, can reveal more about the nature of the dust itself, its production mechanisms, and the processes by which it evolves. 

The evolution of the DTM over cosmic time is a tracer of the history of the interplay between gas and dust in the ISM of galaxies, and its distribution in comparison to metallicity can be used to infer clues about the origin of interstellar dust. If all dust and metals were to be produced in and ejected from stars, one would expect the DTM to remain constant in both time and metallicity (e.g. \citealt{Franco1986}). In models, this is often assumed (e.g. \citealt{Edmunds1998}), especially in the local Universe \citep{Inoue2003}, and a fairly constant dust-to-metals ratio is indeed observed \citep{Issa1990, Watson2011}. At higher redshift, \citet{Zafar2013} found that the DTM in a sample of foreground absorbers to gamma-ray bursts (GRBs) and quasars tends not to vary significantly over a wide range of redshifts, metallicities,  and hydrogen column densities, proposing a universally constant DTM. \citet{Chen2013} find a slow redshift evolution of DTM in lensed galaxies. These findings suggest that most of the dust is produced `instantaneously' in the ejecta of core-collapse supernova (CCSNe), a result supported by recent models by \citet{McKinnon2016}, who find that roughly two-thirds of the dust in MW-like galaxies at $z=0$ is produced in Type II SNe. These authors all use the traditional method of measuring DTM: the extinction, $A_V$, is compared to the equivalent metal column density, $\log N(\mathrm{H}) + $[M/H], where [M/H] is the logged metallicity of the gas (see Eq. \ref{eq:relative}).

Other studies use a different definition of DTM, namely by determining the dust fraction $\mathcal{F}_d$ from the dust depletion (Sect. \ref{sec:depletion}) of metals observed in damped Lyman-$\alpha$ absorbers (DLAs) on sight lines to quasars \citep{Vladilo2004} and GRBs \citep{DeCia2013}. These studies, unlike  those using  $A_V$ as their dust quantifier, claim detections of increasing evolution of the DTM with metallicity. This would suggest that the majority of the dust is formed by growth onto grains in the ISM \citep{Draine2009} rather than simultaneously together with the metals formed in CCSNe and post-AGB star envelopes. \citet{Tchernyshyov2015} use depletion observations in the Small Magellanic Cloud (SMC) to suggest the trend between DTM and metallicity only occurs below a certain metallicity threshold that depends on gas density. \citet{Mattsson2014} provide a comprehensive discussion on the debate from a theoretical standpoint, suggesting that selection effects or statistical fluctuations could  explain the differing observed trends, and \citet{Feldmann2015} attempts to model the observed evolution of dust and metal parameters via production, accretion, destruction, as well as gas  infall and outflow from the galaxy, and also reproduce an evolution of the DTM at low metallicities. \citet{McKinnon2016} include stellar production and accretion along with destruction by SN shocks and winds driven by star formation in models that predict the DTM of MW-like galaxies.

GRBs are useful tools with which to study trends in the DTM in the distant Universe. They are extremely bright, allowing their detection even at very high redshift \citep{Tanvir2009}, and occur in galaxies with a wide range of dust content and metallicities (e.g. \citealt{Fynbo2008,Mannucci2011,Kruehler2015,Cucchiara2015}). GRBs are massive stellar explosions (e.g. \citealt{Galama1998}), the afterglows of which are observed to have featureless synchrotron spectra \citep{Meszaros1997}. This means that any absorption lines or changes to the shape of the spectrum must originate from an absorbing medium between the explosion site and the observer. Typically they manifest themselves in the form of DLAs in the host galaxy of the GRB.
A DLA is defined as an absorbing system with $\log(N(\ion{H}{I}))>20.3$ \citep{Wolfe2005}, and it has been found that a large proportion of GRB afterglow spectra that lie in the redshift range for the Ly-$\alpha$ transition to fall into the atmospheric transmission window ($z>\sim1.7$) do indeed fulfil this criterion (e.g. \citealt{Kruehler2013,Sparre2014,Friis2015}). With such a large pool of neutral gas, the ionization fraction is so small  that the dominant state of the elements used in this analysis is the singly ionized one \citep{Wolfe2005,Viegas1995,Peroux2007}, and the measurements of singly ionized metal species are taken to be representative of the total gas phase abundance of these metals in the DLA. We do commonly detect highly ionized species such as $\ion{C}{IV}$ and $\ion{Si}{IV}$, both often saturated, which might call the above assumption into question, and \citet{Fox2004} do indeed use the ratio [$\ion{C}{IV}/\ion{O}{VI}$] as proportional to the total [C/O]. However, these lines  often show broader velocity structure and/or offsets in central velocity than the low-ionization lines (e.g. \citealt{Fox2007}), suggesting that the gas with a higher ionization state does not trace the same structure as the low-ionization lines. This issue is also addressed in \citet{Ellison2010}, and while they suggest that there may be some ionization corrections below $\log(N(\ion{H}{I}))<21$, they are still low, and only two of the objects in our sample have a neutral hydrogen column density below this value. We thus make no ionization corrections throughout the paper, and take the low-ionization abundances to be representative. 

In this paper, we present spectral analysis of five previously unpublished GRBs, and we combine them with 14 more GRB-DLAs from the literature, all but three of which have mid- to high-resolution spectroscopy. We compute dust depletion curves using all of the available metals, which we then use to calculate average DTM values, and investigate their relation with metallicity and redshift in order to investigate the evolution of DTM.

The structure of this paper is as follows. In Sect. \ref{sec:depletion} we describe the background and updated methods available to parameterize dust-depletion. The initial sample is presented in Sect. \ref{sec:sample}. In Sect. \ref{sec:method} we introduce our method of fitting for depletions in GRB-DLAs, and in Sect. \ref{sec:results} we present the results; we discuss the results in Sect. \ref{sec:discussion} and conclude in Sect. \ref{sec:conc}.
Throughout the paper we assume the solar abundances from \citet{Asplund2009}. 


\section{Dust depletion\label{sec:depletion}}

Using spectroscopy, it is possible to measure the column density of ISM constituents through absorption lines. However, what is achieved here is a measure of the gas phase abundance of that element, as any metal atoms in the dust grains do not contribute to the observed absorption. The difference between the observed column and intrinsic, total column density of metal $X$ is referred to as dust depletion
\begin{equation}
\label{eq:delta_def}
  \delta_{X} = [X/\mathrm{H}]_{\mathrm{obs}} - [X/\mathrm{H}]_{\mathrm{in}}\,,
\end{equation}
where a greater amount of element X is expected to be depleted onto dust grains with increasing negative values of $\delta_X$. We use the standard relative abundance notation,
\begin{equation}
\label{eq:relative}
 [X/Y]= \frac{\log{N(X)}}{\log{N(Y)}} - \frac{\log{N(X)}_{\mathrm{ref}}}{\log{N(Y)}_{\mathrm{ref}}} \,.
\end{equation}

To calculate the amount of depletion, we need to know two things: the observed gas-phase abundance of each element, and the total intrinsic (gas + dust) abundance. The observed column densities are obtained from the GRB afterglow spectrum, but the intrinsic values are harder to come by, as we do not know {a priori} the total column density of a metal in both the gas and solid phase ($\textnormal{i.e.}$ dust). 
It was shown by \citet{Savage1996} that different elements deplete onto dust at different rates. Some elements, such as Fe and Ni, deplete rapidly and are known as refractory elements. Others, such as Zn, P, S, and Si, are almost always entirely in the gas phase and are denoted as volatile. The measured abundance of these volatile elements  in this formulation are taken to be a good indicator of the metallicity of the system, and the difference in the relative abundances of a volatile and a refractory element, such as the ratio [Zn/Fe], is thus a basic quantifier of depletion. 

\citet{Savage1996} measured the depletions of several elements towards a set of different sight lines in the MW, from the dust-poor warm halo (WH) clouds, increasing in dust content to warm disk+ halo (WDH), warm disk (WD), and finally heavily dusty cold disk (CD) clouds. They reported typical depletion levels of each element for each cloud (Fig. 5 and Table 6 in \citealt{Savage1996}).
When investigating depletion in a DLA,   fitting each of the MW depletion patterns to the observed relative abundances can be attempted. Since the metallicity and dust-to-metals ratio in the DLA are likely to be different to the MW, they are left as free parameters;  the relative abundances expected from the model are adjusted until they best match those of the observed abundances, a method described by \citet{Savaglio2001} and \citet{Savaglio2003}.
Although these methods provide a basis for depletion studies, it is often found that GRB-DLAs tend not to follow any Local Group depletion patterns particularly well (e.g. \citealt{DElia2014, Friis2015}).

Based on the concept of \citet{Savage1996}, a more continuous determination of depletion was introduced by \citet{Jenkins2009}, based on depletions observed in 17 elements towards stars along 243 MW sight lines. It was found that all elements deplete in a linear fashion, such that the rate of depletion of an element $X$, $\delta_X $, can be given as  
\begin{equation}
\delta_{X} = B_{X} + A_{X}\left(F_{*}-z_{X}\right)\,,
\label{f*_def}
\end{equation}
where $A_{X}$ is the depletion slope, and  $ B_{X}$ and $z_{X}$ are constant offsets. This formulation implies that the difference between the depletion of any two elements should depend only on the value $F_*$, the depletion strength factor of the environment. That is, the relative abundances between any set of two or more elements in a single sight line can only be described by one unique value of $F_*$, which is then a powerful tool that can be used to  calculate an overall, average DTM using multiple elements.
   \begin{figure}
   
   \centering
   \includegraphics[width=\hsize]{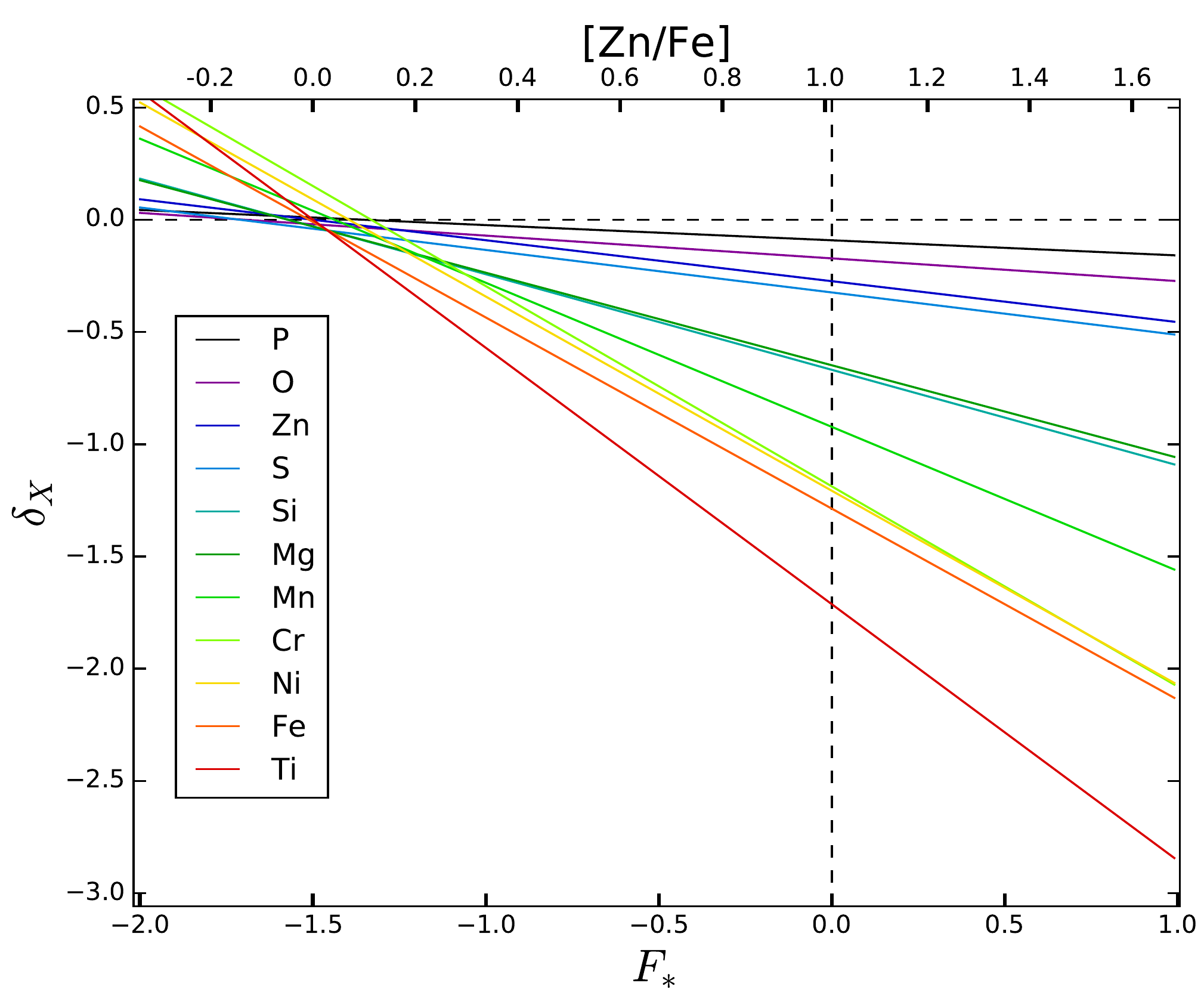}
   
   \caption{\label{fig:f*} Depletion $\delta_X$
                is plotted as a function of 
               $F_*$. The slope is defined by the parameter $B$, and the point of zero-depletion by $A$. The elements are listed in order of depletion strength.}
   \end{figure}
In the \citet{Jenkins2009} formulation, $F_{*} = 0$ is given to those sight lines where the least depletion was observed, with $F_{*} = 1$ the value for the most depleted MW systems. The $F_*$ method is also applied to the abundances in the \citet{Savage1996} models, with the WH clouds having $F_{*} = -0.28$, whereas $F_{*} = 0.90$ in the dusty CD environment.

A study of the Magellanic Clouds (MCs) was presented by \citet{Tchernyshyov2015} who used the $F_*$ method to ascertain whether depletion slopes and offsets followed the MW trends from \citet{Jenkins2009}. Here they combined the offsets $A_X$ and $z_X$ into one offset, $\delta_0$, which simply describes the level of depletion at $F_* = 0$. Compared to the MW, they found differences in the offsets for the elements P, Zn, Si, Cr, and Fe, and that $\delta_{0}$ decreases with metallicity of the sight line. Depletion slope $A_{X}$, however, tends to stay constant. The main interpretation of this is that at lower metallicities, a higher $F_{*}$ is needed before metals begin to form dust, although this is based on a fairly small sample covering only the metallicities of the Small and Large Magellanic Clouds.

Recently, the most comprehensive study yet of depletion patterns down to low metallicity and dust content has been conducted by \citet{DeCia2016}, detailing depletion sequences of nine elements (P, O, Zn, S, Si, Mg, Mn, Cr, Fe) in 70 QSO-DLAs at redshifts of 1.7-4, and metallicities from the solar value down to $1/100 Z_{\sun}$. 
They use a very similar method to \citet{Jenkins2009}:  the depletion of each element in each DLA is plotted against [Zn/Fe], used here as the dust quantifier instead of Jenkins' $F_*$, although the two parameters are directly and linearly linked. The Zn is used as a proxy for an undepleted element, but in reality this is not the case. It is corrected for its own depletion by comparing the [Zn/H] and [Zn/Fe] values in the MW. For a measured [Zn/Fe] in a QSO sightline, \citet{DeCia2016} interpolate the MW [Zn/Fe] - [Zn/H] relation to find the corresponding Zn depletion. This missing Zn is then added to give a corrected [Zn/Fe] to be used as the final depletion strength indicator for that sight line.
To add continuity to the depletion sequences, these QSO data are supplemented with Jenkins' data from the MW, and it can be seen that the high-$z$ data show the same behaviour as -- and are a simple continuation of -- those from the MW. For each element, \citet{DeCia2016} then use linear regression techniques to fit the depletion sequences and calculate two parameters: $A$, the depletion at [Zn/Fe] = 0, and $B$, the depletion slope. These depletion sequences are shown in Fig. \ref{fig:f*}. There are distinct clusters of elements: P, O, Zn, and S, which are volatile elements;  Cr, Ni, and Fe, which are refractory elements;  Si, Mg, and Mn, which lie  between the volatile and refractory elements; and Ti, which seems to lie distinctly below all of the others.

From Fig. \ref{fig:f*} it can be seen that at  $F_{*} = 0$ there is still significant depletion ($\sim1$ dex), especially in the refractory elements, which shows that even the least depleted MW clouds are more dusty than those in low-metallicity QSO-DLAs. 
The parameter $A$ can best be seen as the offset in the relative abundance of the element at the point of no iron depletion, [Zn/Fe]=0. Since they are all very small, the point at which the depletion starts (i.e. $\delta_{X} = 0$ ) is very similar for each element, suggesting that all elements begin forming dust at some distinct threshold in temperature or density.

The analysis from \citet{DeCia2016} shows that the slopes are linear and well determined in the whole range from low metallicity and low dust content right through to the dustiest MW clouds, which we believe is strong evidence that the depletion mechanism works in the same way in all environments, and can indeed always be quantified by a single depletion strength factor.  For this reason, we use the slopes from \citet{DeCia2016} for our depletion analysis of the typically low-dust, low-metal GRB-DLA environments in order to calculate DTMs and metallicities, which we describe in Section \ref{sec:method}.

\section{Sample and data reduction\label{sec:sample}}

\begin{sidewaystable*}
\caption{Hydrogen and metal abundances in our sample of GRB afterglows. Units of column density are $\log$(cm$^{-2}$).}             
\label{table:1}      
\centering
\setlength{\tabcolsep}{2pt}          
\begin{tabular}{c c c c c c c c c c c c c c c c  }     
\hline\hline       
GRB & $z$ & $ \log N$(Mg) &  $ \log N$(Si) & $ \log N$(P)& $ \log N$(Ti)&  $ \log N$(Cr) &  $ \log N$(Mn) &  $ \log N$(Fe)&  $ \log N$(Ni)&  $ \log N$(Zn)& $ \log N$(S)\\ 
\hline                    
   \object{000926} &  2.04     & - & $16.47\pm0.05^1$ & - & - & $14.34\pm 0.05^1$ & - & $15.60\pm0.20^1$ & - &$13.82\pm0.05^1$ \\
   \object{050401} &  2.9       & - & $16.5\pm0.4^2$ & - & - & $14.6\pm0.2^2$ & - & $16.0\pm0.2^2$ & - & $14.3\pm0.3^2$ & -\\
   \object{050730} &  3.9672 &<$16.08^4 $& $15.47\pm0.03^4$ & - & - & - & - & $15.49\pm0.03^3$ & $13.69\pm0.02^3$ & - & $15.11\pm0.04^3$\\
   \object{050820A} &  2.6145 & $15.86\pm0.05^4$ & - & $13.64\pm0.04^4$ & - & $13.33\pm0.02^4$ & - & $14.82\pm0.12^4$ & $13.69\pm0.04^4$ & $12.96\pm0.02^4$ & $15.57\pm0.04^4$ \\
   \object{070802} &  2.45     &  - & $16.00\pm0.32^5$ & - & - & $14.04\pm0.40^5$ & $13.69\pm0.18^5$ & $16.16\pm0.18^5$ & $14.89\pm0.28^5$ & $13.67\pm0.65^5$ & - \\
   \object{081008} &  1.9685 &  - & $15.75\pm0.04^6$ & - & - & $13.83\pm0.03^6$ & - & $15.42\pm0.04^6$ & $13.74\pm0.07^6$ & $13.15\pm0.04^6$ & - \\
   \object{090323} &  3.567   & - & $15.80\pm0.05^7$ & - & - & $-$ & - & $15.00\pm0.05^7 $& $14.76\pm0.04^7$ & $13.57\pm0.04^7$ & $15.80\pm0.02^7$       \\
   \object{090809} &  2.737   & $>16.80^8$ & $16.15\pm0.07^8$ & - & - & - & $13.75\pm0.12^8$ & $15.75\pm0.07^8$ & $14.40\pm0.07^8$ & $13.70\pm0.25^8$ & -     \\
   \object{090926A} &  2.1071 &  $>14.11^9$ & $14.80\pm0.08^9$ & - & - & - & - & $14.86\pm0.09^9$ & $13.92\pm0.13^9$ & - & $14.89\pm0.10^9$      \\
   \object{100219A} &  4.667   &  $>13.72^{10}$ & $15.25\pm0.25^{10}$ & - & - & - & - & $14.73\pm0.11^{10}$ & - & - & $15.25\pm0.15^{10}$ \\ 
   \object{111008A} &  4.99     &  - & $>15.86^{11} $& - & - & $14.17\pm0.11^{11}$ & $13.72\pm0.08^{11}$ & $16.05\pm0.05^{11}$ & $14.89\pm0.18^{11}$ & $13.28\pm0.21^{11}$ & $15.71\pm0.09^{11}$   \\
   \object{120119A} &  1.7285 & - & $16.67\pm{0.35}^{12}$ & - & $13.18\pm{0.25}^{12}$ & $14.21\pm{0.20}^{12}$ & $13.99\pm{0.20}^{12}$ & $15.95\pm{0.25}^{12}$ & $14.77\pm{0.17}^{12}$ & $14.04\pm{0.25}^{12}$  & -      \\
   \object{120327A} &  2.8145 &  $16.34\pm0.02^{13}$ & $16.36\pm0.03^{13}$ & $14.19\pm0.04^{13}$ & - & $14.17\pm0.02^{13}$ & - & $15.78\pm0.09^{13}$ & $14.61\pm0.09^{13}$ & $13.40\pm0.04^{13}$ & $15.74\pm0.02^{13}$       \\
   \object{120716A} &  2.487   & - & $16.48\pm{0.45}^{12}$ & - & - & $14.20\pm{0.26}^{12}$ & $14.02\pm{0.26}^{12}$ & $15.65\pm{0.45}^{12}$ & $14.42\pm{0.22}^{12}$ & $13.91\pm{0.32}^{12}$ & - \\
   \object{120815A} &  2.358   &  - & $\geq 16.34^{14}$ & - & - & $13.75\pm0.06^{14}$ & $13.26\pm0.05^{14}$ & $15.29\pm0.05^{14}$ & $14.19\pm0.05^{14}$ & $13.47\pm0.06^{14}$ & $\leq16.22^{14}$      \\
   \object{120909A} &  3.929   &  $\leq16.55^{12}$ & $16.22\pm{0.32}^{12}$ & - & - & - & - & $15.20\pm{0.18}^{12}$ & $14.36\pm{0.13}^{12}$ & $13.55\pm{0.32}^{12}$ & $16.10\pm{0.18}^{12}$ \\
   \object{121024A} &  2.30     &  - & $>16.35^{15}$& - & - & $14.18\pm0.03^{15}$ & $13.74\pm0.03^{15}$ & $15.82\pm0.05^{15}$ & $14.70\pm0.06^{15}$ & $13.74\pm0.03^{15}$ & $>15.90^{15}$\\
   \object{130408A} &  3.7579 &  $16.01\pm0.21^{12}$ & $15.95\pm{0.22}^{12}$ & - & $12.77\pm{0.22}^{12}$ & $13.81\pm{0.13}^{12}$  & $13.19\pm{0.13}^{12}$ & $15.52\pm{0.11}^{12}$ & $14.15\pm{0.08}^{12}$ & $12.87\pm{0.16}^{12}$ & $15.78\pm0.18^{12}$\\
   \object{141028A} &  2.333   &  $15.28\pm0.27^{12}$ & $14.82\pm{0.33}^{12}$ & - & - & $12.98\pm{0.33}^{12}$ &$<13.29^{12}$& $14.23\pm{0.21}^{12}$ & $13.34\pm{0.33}^{12}$ & $12.38\pm{0.33}^{12}$ & -     \\
\hline        
       
\end{tabular}
\tablebib{(1)~\citet{Savaglio2003}; (2)~\citet{Watson2006}; (3)~\citet{Ledoux2009}; (4)~\citet{Prochaska2007b}; (5)~\citet{Eliasdottir2009}; (6)~\citet{DElia2011}; (7)~\citet{Savaglio2012}; (8)~\citet{Skuladottir2010}; (9)~\citet{DElia2010}; (10)~\citet{Thoene2013}; (11)~\citet{Sparre2014}; (12)~\textit{this work}; (13)~\citet{DElia2014}; (14)~\citet{Kruehler2013}; (15)~\citet{Friis2015}}   
\end{sidewaystable*}

In order to select our sample, we require GRBs with mid- to high-resolution spectroscopy as this allows us to resolve relatively well the velocity structure of the absorber, thus limiting the effect of unidentified saturation in the lines. The spectrum must also cover the Ly-$\alpha$ line, such that we can verify that it is a DLA, and thus regard any ionization correction as negligible. Finally, we require that the spectrum includes unsaturated detections of at least four singly ionized metals in order to increase the precision on a measurement of $F_*$.\footnote{We select GRBs up to the end of 2014. There are more recent GRBs that also pass the selection criteria, but it is beyond the scope of this paper to keep adding to the sample.}\\
 A state-of-the-art example of an instrument that produces such spectra is X-shooter (Vernet et al., 2011), mounted at ESO's VLT at Cerro Paranal, Chile. X-shooter operates simultaneously in three spectral arms, namely the bands UVB (3000-5500 \AA), VIS (5500-10000 \AA), and NIR (10000-25000 \AA), thus allowing absorption line metallicity measurements from redshifts $\gtrsim 1.8$, and providing a wide spectral range. It operates at a resolving power of around $R = 8000$, depending on the arm, slit used, and atmospheric conditions. 
Twelve GRBs observed with X-shooter pass our selection criteria. Seven of these, GRBs \object{090809}, \object{090926A}, \object{100219A}, \object{111008A}, \object{120327A}, \object{120815A,} and \object{121024A} (references in Table \ref{table:1}), have already been published in the literature, and we include them in the sample. In this paper we present the analysis for the remaining five\footnote{Spectra of GRBs 120716A, 120909A, and 130408A have been analysed by \citet{Cucchiara2015}, but that work presents column density measurements for only Ly-$\alpha$ and one other element, and does not show line fits and velocity components, as we do here.}: 120119A, 120716A, 120909A, 130408A, and 141028A (see Sect. \ref{subsec:fitting}).
A further two GRBs observed with UVES pass the selection criteria (\object{050730} and \object{081008}, both taken from the literature), and an additional two GRBs observed with Keck HIRES and ESI have data published and are therefore also included in sample (\object{050820A} and \object{000926,} respectively). Unpublished GRBs with Keck high-resolution spectral data cannot be included in the sample since data are not public. In order to help populate the sample with dust-rich sight lines, we add three low-resolution (VLT/FORS) spectra of GRBs \object{050401}, \object{070802,} and \object{090323}, which have high $N(\mathrm{H})$, high $A_V$, and high metallicity, respectively. However, given the uncertainty in the derived column densities for these GRBs, we distinguish them from the rest of the sample when presenting our results. The final sample of 19 GRBs along with references is presented in Table \ref{table:1}; the $\ion{H}{I}$ measurements are given in Table \ref{table:results}.

\subsection{New GRB Spectra \label{subsec:fitting}}

For the spectra of GRBs \object{120119A}, \object{120716A}, \object{120909A}, \object{130408A,} and \object{141028A} we perform our own analysis on the spectra obtained from X-shooter. The general method used to reduce the raw spectra is based on the standard X-shooter pipeline \citep{Goldoni2006,Modigliani2010}, which we modify in accordance with the procedures outlined in \citet{Kruehler2015}, including a correction for telluric absorption using the Molecfit software \citep{Smette2015}. To normalize, we select points on the continuum unaffected by absorption lines, and fit a spline function.
Owing to good seeing, the measured resolving power is often larger than the value  determined from arc lamp exposures (see e.g. \citealt{Kruehler2013}). We therefore follow the standard procedure, which is to measure the velocity resolution from unsaturated, single telluric lines. Since there are no telluric lines in the UVB, we use the resolution of those measured in the other two arms, calculating the resolution in the UVB using known conversion factors in line with \citet{Fynbo2011}.
We perform Voigt-profile fits on the absorption lines using the line-fitting software \texttt{VPFit v.10.2}\footnote{VPFit: http://www.ast.cam.ac.uk/ rfc/vpfit.htm}. At resolutions typically around 30 km s$^{-1}$, we often resolve multiple velocity components, each with distinct $b$-parameters. We determine the nature of any such components by fitting singly ionized, unsaturated, and unblended transitions;  e.g. $\ion{Fe}{II}$  $\lambda1611$, $\ion{Ni}{II}$ $\lambda\lambda$1751, 1741, $\ion{Si}{II}$ $\lambda1808$, and $\ion{Mn}{II}$  $\lambda2606$ are often useful transitions. We then fix redshift, $z$, and broadening parameter, $b~\mathrm{km~s}^{-1}$, for each velocity component across all species, leaving column density $N$ cm$^{-2}$ as the free parameter. We present the resulting column densities in Table \ref{table:1}, and present a selection of line fits for each GRB in Appendix A. 

\subsubsection{Hidden saturation\label{subsubsec:hidden saturation}}

An issue that commonly plagues absorption line astronomy, particularly at mid to low resolution and poor signal-to-noise ratio (S/N), is that of hidden saturation. This occurs when a line that is in reality marginally saturated is smoothed out by the instrument resolution and noise to appear unsaturated in the actual observed spectrum. The pitfalls of this have been well documented by e.g. \citet{Prochaska2006,Rafelski2012,Jorgenson2013} and \citet{Cucchiara2015}, who typically ignore any lines that show any sign that they could be saturated.
However, because the basis of our analysis is the use of multiple species to constrain the metallicity and dust depletion, we require measurements of as many lines as possible in the often low S/N GRB spectra. 

\begin{figure}
   \centering
 
 \includegraphics[width=\hsize]{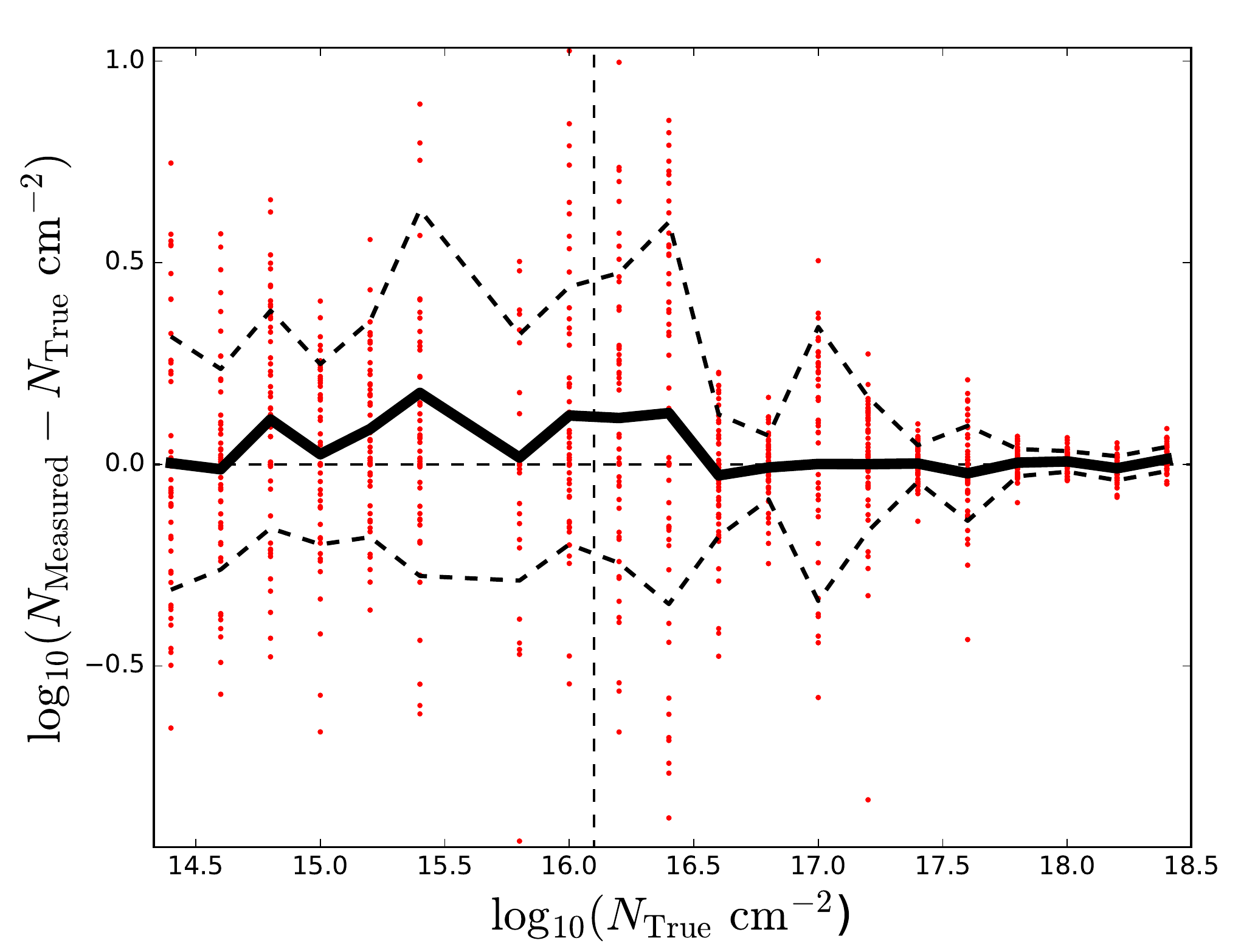} 
   
   \caption{Difference between the measured column density of $\ion{Si}{II}$ $\lambda1526$ after convolving with the X-shooter resolution and adding  noise at S/N = 7.5, similar to the lowest quality of our data. The thick black line shows the mean residual, while the dashed curves show the 1$\sigma$ level. The vertical dashed line marks the point at which this line saturates in the original spectrum. }
              \label{fig:simuls}
    \end{figure}

To investigate and quantify the effect this has on our results we run a set of simulations, similar to those of \citet{Jorgenson2013} who used the apparent optical depth method (AODM). Initially we take the worst-case scenario, and assume the lowest S/N and highest resolution in our GRB sample spectra, which corresponds to 7.5 and 35.0 km s$^{-1}$, respectively, in the case of \object{GRB 120716A}. We then simulate a set of absorption lines of one transition, namely $\ion{Si}{II}$ $\lambda1526$, at a common $b$-parameter but with increasing column density so as to straddle the theoretical point at which that line saturates. We choose a $b$-parameter of 10 km s$^{-1}$ as this represents the smallest value typically seen with higher resolution instruments, with smaller components typically blended \citep{Jorgenson2013}. We create lines starting from unsaturated column densities of 14.4 cm$^{-2}$ up to heavily saturated column densities of 18.4 cm$^{-2}$, and convolve with the X-shooter resolution measured in \object{GRB 120716A}. We add Gaussian noise at a S/N of 7.5 and use \texttt{VPFit} to fit a Voigt profile and measure the column density. The results for 100 trials, with different noise added each time, are plotted in Fig. \ref{fig:simuls}. The thick black curve shows that on average the measured column densities are always within 0.2 dex of the true value, with a standard deviation around 0.4 dex. Above 16.6 cm$^{-2}$ there is on average no deviation, and above 14.4 cm$^{-1}$ the data points have a standard deviation lower than 0.1 dex, reflecting the fitting of the damping wings. 

From these simulations, we find that there is only a small loss of accuracy of our column density measurements when lines are saturated, although we do continue to avoid using strongly saturated lines, especially when several components are evident. 
We also find that the uncertainty on the column densities provided by \texttt{VPFit} is often an underestimate compared to that from our simulations. To find the uncertainty for all of our measurements, including those on lines with potential hidden saturation, or evident but mild saturation, we thus conduct a second round of simulations. For each GRB, we take the S/N and resolution measured from the spectrum and simulate the line with the above method 100 times, but this time for only one unsaturated column density of 15.8 cm$^{-2}$. We take the standard deviation of our measurements as the error on a single line measurement for that GRB. For species with multiple lines used, we add the errors from each line in quadrature.

\section{Method\label{sec:method}}

\subsection{Depletion model fitting\label{subsec:modelling}}

Using the concept of linear depletion sequences and $F_*$ from \citet{Jenkins2009}, \citet{DeCia2016} use the observable [Zn/Fe] as their dust indicator. It is possible to directly translate between the two with the relation $F_* = 1.48 \times [\mathrm{Zn/Fe}]-1.50$. To avoid confusion between [Fe/Zn] as an intrinsic measured value, and ``adjusted'' [Fe/Zn] as a dust indicator, we use $F_*$ as our depletion strength factor. For a given metallicity and dust depletion strength, the relative abundance in element $X$ that  we expect to measure according
to the model is thus
\begin{equation}
 [X/\mathrm{H}]_{\mathrm{exp}} = \delta_X + [\mathrm{M/H}] =  A + B \frac{\left(F_*-1.50\right)}{1.48} + [\mathrm{M/H}]\,,
 \label{xhmod}
\end{equation}
where [M/H] is the metallicity of the system, and $A$ and $B$ are the updated linear depletion parameters from \citet{DeCia2016}. 
In a similar way to \citet{Savaglio2003}, we vary $F_*$ and [M/H] to minimize the $\chi^2$ parameter between the observed abundances and those expected from the model, thus achieving a best fit depletion strength and metallicity. The 1 $\sigma$ errors on the parameters are calculated for a single parameter of freedom. 
Typically, volatile elements such as Zn, P, S, and Si are used as metallicity tracers, often left uncorrected for dust (e.g. \citealt{Cucchiara2015}). Our method goes one step further: with the large spectral range of X-shooter, we use the information from all of the possible species to fit for dust depletion and thus retrieve a metallicity. We are therefore less sensitive to the pitfalls of only using a particular metallicity tracer.

The errors on $F_*$ and [M/H] are dependent on the errors on the column densities of $\ion{H}{I}$, each individual element, and on the number of elements included in the dust depletion curve fits. Therefore, when only four elements are available or when the column density measurements are not tightly constrained, the uncertainty on the metallicity and $F_*$ can be quite large. The 2D chi-squared contour plots are provided in Appendix B. The shape of these plots show elongated confidence regions, which can be seen as a degeneracy between the two parameters. In most cases $F_*$ and [M/H] are still well constrained, and when this is not the case it is reflected in large errors, for example in GRB 050401.

Unfortunately, \citet{DeCia2016} did not publish measurements for Ni or Ti.  Nickel in particular is measured in the majority of our spectra; it is  a strongly refractory element with depletion properties similar to Fe, and provides valuable information on the dust content of the DLA. To calculate Zn depletion, \citet{DeCia2016} use the \citet{Jenkins2009} MW data and a least-squares method to fit the slope between [Zn/Fe] and [Zn/H]. We use the same technique with Ni and Ti, using column densities from \citet{Jenkins2009} and orthogonal distance regression to linearly fit the data and retrieve $A$ and $B$ parameters. We have seen that the slopes measured down to low dust content are compatible with those measured only in the Galaxy to within the uncertainties, and as such we trust that our Ni and Ti $A$ and $B$ values also follow this trend, and provide model values consistent with those for the other elements.

\subsubsection{Nucleosynthesis \label{subsec:ref_abundances}}

Dust depletion analysis relies on the difference between an observed and an expected, intrinsic abundance for each metal. We use the solar abundances from \citet{Asplund2009} as our reference. This could, however, lead to errors in the depletion calculation. Similar to the composition of dust grains, it is perfectly logical to assume that intrinsic abundances in a qqq high-redshift DLA are somewhat different to those observed in the Sun. 

One of the most common nucleosynthetic effects at high redshift and low metallicity is an overabundance of $\alpha$-elements such as O, Si, S and Mg, in comparison to Fe, often denoted by the factor [$\alpha/$Fe.] The \citet{DeCia2016} depletion patterns have been corrected for these effects, and we adopt their method of applying corrections in our work. In short, this involves applying the observed trend between [Zn/Fe] and [Zn/H] to use [Zn/Fe] as a basic proxy to estimate metallicity. We then use conversions provided by \citet{DeCia2016} which show the nucleosynthetic correction to the abundances at that [Zn/Fe], based upon their observations alongside those by \citet{Lambert1987,McWilliam1997,Wheeler1989}, for $\alpha$-enhancement, and \citet{Wheeler1989,Mishenina2015,Battistini2015} for Mn.

\citet{Vladilo2011} calculate the reference abundances in a more theoretical way, using galaxy evolution models to predict metal abundances. It would be interesting to see how such an approach affected our results, but such an analysis is beyond the scope of this paper.

   \begin{figure*}
  
            {\includegraphics[width=19cm]{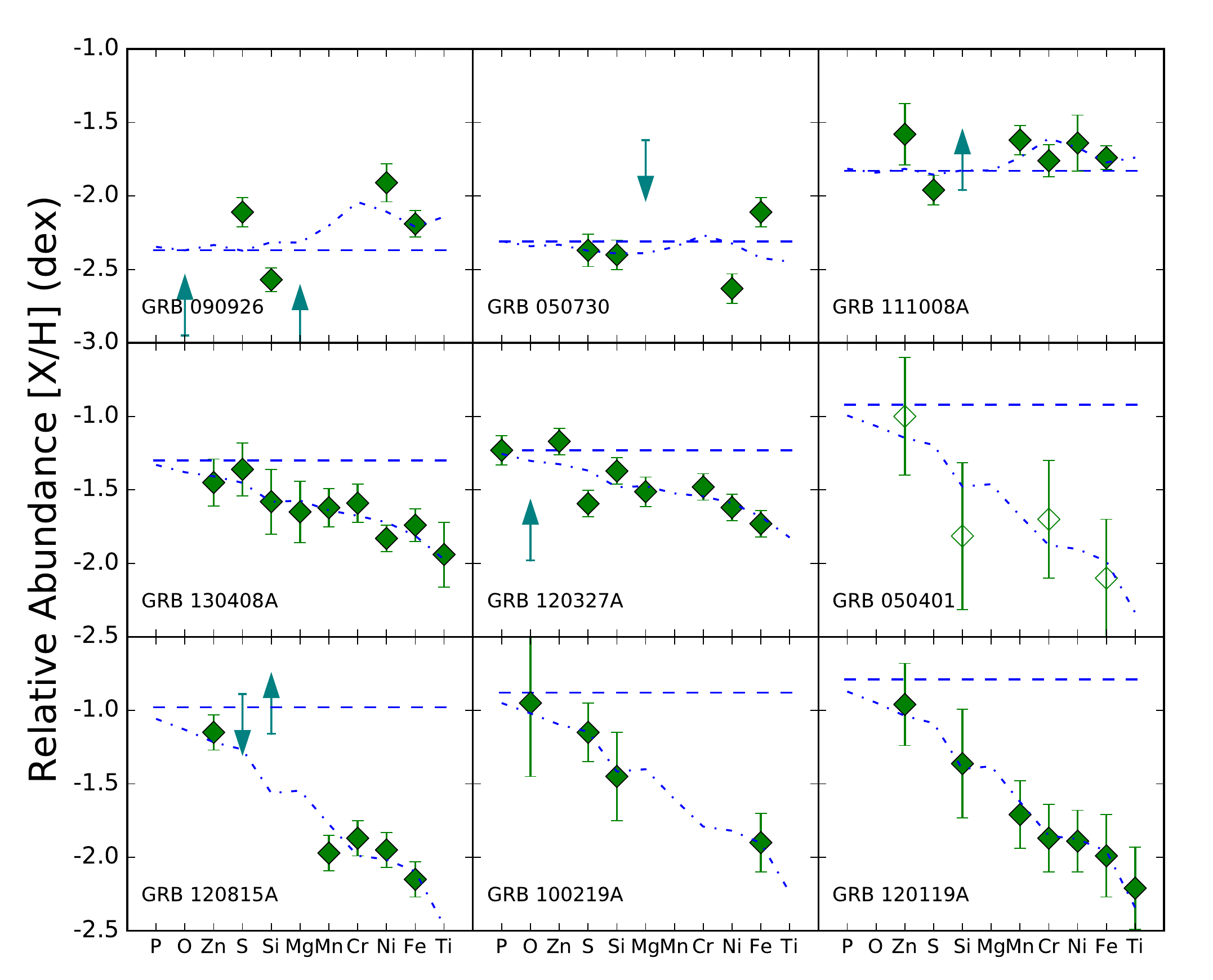}
      \caption{Dust depletion fits for the nine lowest metallicity GRB-DLAs in our sample. The diamond points are the observed relative abundances, the dot-dashed line follows the expected depletion at a strength of the best fit $F_*$, and the dashed line represents the best fit metallicity. Unfilled markers represent low-resolution spectral data (we continue this in all following plots), and squares with up (down) arrows represent lower (upper) limits.
              }
         \label{fig:depletions1}}
   \end{figure*}
  \begin{figure*}
  
            {\includegraphics[width=19cm]{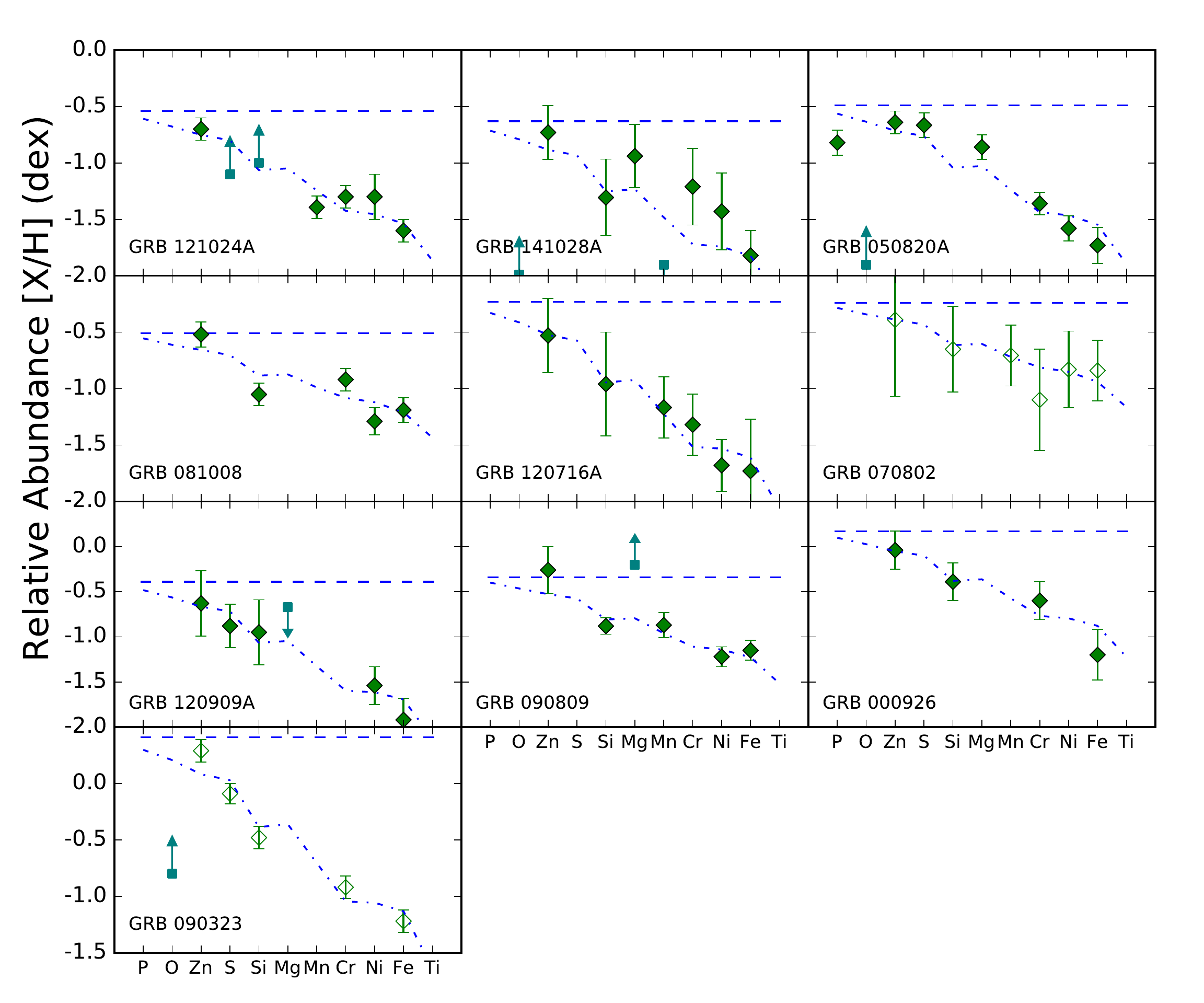}
      \caption{A continuation of Fig. \ref{fig:depletions1} for the ten GRB-DLAs with highest metallicity.
              }
         \label{fig:depletions2}}
   \end{figure*}
 \begin{table*}
\caption{Results of the fitting of the $F_*$ depletion model to metal column densities from 19 GRB-DLAs. Given uncertainties are at the 1$\sigma$ level.}             
\label{table:results}      
\centering          
\begin{tabular}{c c c c c c c c c  c }     
\hline    
                       
GRB & $A_V$ (mag) &$ \log (N$(H) cm$^{-2})$& $F_*$ & [M/H] & \DTM\ \\ 
\hline                    
  
   \object{000926} &   $0.38\pm0.05^{1}$  & $21.30\pm 0.20$  & $ -0.28\pm0.18  $ & $0.17\pm0.34$   & $ 0.76\pm0.07$ \\
   \object{050401} &   $0.45\pm0.035^{2}$  & $22.60\pm 0.30$  &  $ -0.26\pm0.31  $ & $-0.92\pm0.68$   & $ 0.76\pm0.09$\\
   \object{050730} &   $\leq 0.17 ^{2}$        & $22.10\pm 0.10$  &  $ -1.38\pm0.07  $ & $-2.31\pm0.18$ &$ 0.19\pm0.02$\\
   \object{050820A} &   $0.27\pm0.05^{3}$   & $21.05\pm 0.10$  &  $ -0.27\pm0.07  $ & $-0.49\pm0.10$&$ 0.76\pm0.06$\\
   \object{070802} &   $1.23\pm0.05^{4}$  & $21.50\pm 0.20$  &  $-0.69\pm0.18  $ & $-0.24\pm0.80$ &   $ 0.59\pm0.06$\\
   \object{081008} &   $\leq 0.08^{4}$         & $21.11\pm 0.10$  &  $-0.69 \pm 0.06 $ & $-0.51\pm0.17$ &  $ 0.59\pm0.04$\\
   \object{090323} &   $0.10\pm0.04^{2}$     & $20.72\pm 0.09$  & $0.30 \pm0.07  $ &  $0.41\pm0.11$   & $ 0.95\pm0.07$ \\
   \object{090809} &   $0.11\pm0.04^{5}$  & $21.40\pm 0.08$  &  $ -0.47\pm0.07  $ & $-0.34\pm0.25$  &$ 0.68\pm0.05$\\
   \object{090926A} &   $\leq 0.01^{6} $        & $21.60\pm 0.07$  & $-1.7 \pm 0.07 $ &  $-2.37\pm0.16$ &$ 0.00\pm0.01$ \\
   \object{100219A} &   $0.15\pm0.03^{7}$  & $21.14\pm 0.15$  &  $-0.31\pm0.22  $ & $-0.88\pm0.33$ &$ 0.74\pm0.07$\\
   \object{111008A} &   $0.10\pm0.05^{7}$  & $22.30\pm 0.06$  &  $ -1.58\pm 0.06 $ & $-1.83\pm0.16$ &        $ 0.05\pm0.01$\\
   \object{120119A} &   $1.06\pm0.02^{8}$  & $22.44\pm 0.12$  &  $-0.14 \pm 0.11 $ & $-0.79\pm0.42 $ & $ 0.80\pm0.06$\\
   \object{120327A} &   $\leq 0.02^{8}$        & $22.01\pm 0.09$  &  $ -0.98\pm 0.05$                            & $-1.23\pm0.08$                               &  $ 0.44\pm0.03$\\
   \object{120716A} &   $0.30\pm0.15^{5}$  & $21.88\pm 0.08$  &  $  0.11\pm0.16 $ & $-0.23\pm0.55  $ &   $ 0.88\pm 0.08$\\
   \object{120815A} &   $0.08\pm0.02^{8}$  & $21.95\pm 0.10$  &  $ -0.20 \pm 0.07$                        & $-0.98\pm0.22$                               &   $ 0.78\pm0.06$\\
   \object{120909A} &  $0.16\pm0.04^{8}$   & $21.61\pm 0.06$  &  $0.01\pm0.17   $ & $-0.46\pm0.36  $ & $ 0.85\pm0.07$\\
   \object{121024A} &  $0.21\pm0.03^{8}$   & $21.85\pm 0.15$  &  $-0.34 \pm 0.06 $                  & $-0.54\pm0.18$                                &  $ 0.73\pm0.05$\\
           \object{130408A} &  $0.22\pm0.03^{8}$   & $21.76\pm 0.03$  &  $ -0.91\pm0.06   $ & $-1.30\pm0.19  $ &   $ 0.48\pm0.04$\\
   \object{141028A} &  $0.13\pm0.09^{5}$   & $20.55\pm 0.07$  &  $ -0.18\pm0.17    $ & $-0.50\pm0.38  $ &$ 0.79\pm0.07 $\\

\hline                        
\end{tabular}
\\
\tablebib{(1) \citet{Starling2007}; (2) \citet{Schady2011}; (3) \citet{Schady2012}; (4) \citet{Greiner2011}; (5) \textit{this work}; (6) \citet{Rau2010}; (7) Bolmer et al. \textit{in prep}; (8) Greiner et al. \textit{in prep}   }.
\end{table*}

\subsection{Dust-to-Metals Ratio}

Rather than comparing dust and metal quantities measured by different means (e.g. \citealt{Zafar2013}), or by using only one refractory element to trace the dust, we use the depletion strength factor $F_*$ along with our best fit [M/H] and measured $N$(H) to calculate the total column densities in dust phase for all 11 elements, including those not measured in the spectrum: 
\begin{equation}
\label{eq:X_dust_col}
N\left(\mathrm{X}\right)_{\mathrm{dust}} \approx N\left(\mathrm{H}\right) 10^{[X/\mathrm{H}]_{\sun}} 10^{[\mathrm{M/H}]}\left(1-10^{\delta_X}\right) \mathrm{cm}^{-2}\,,
\end{equation}
where $\delta_X$ is the depletion in element X as calculated from the best fit $F_*$.
We can then sum over the elements to find the total dust column density in terms of atoms in the dust phase per cm$^{2}$.
\begin{equation}
\label{eq:dust_col}
N\left(\mathrm{dust}\right) \approx N\left(\mathrm{H}\right) \frac{Z}{Z_{\sun}} \sum\limits_X 10^{[X/\mathrm{H}]_{\sun}}\left(1-10^{\delta_X}\right) \mathrm{cm}^{-2}\,,
\end{equation}
and similarly for the total metal column:
\begin{equation}
\label{eq:met_col}
N\left(\mathrm{metals}\right) \approx N\left(\mathrm{H}\right) \frac{Z}{Z_{\sun}} \sum\limits_X 10^{[X/\mathrm{H}]_{\sun}} ~\mathrm{cm}^{-2}\,,
\end{equation}
with $Z/Z_{\sun} = 10^{\mathrm{[M/H]}}$. We then take the ratio between the dust and total metal column densities, to find a dust-to-metals ratio for the DLA. We can see that metallicity and $N(\mathrm{H})$ cancel out, such that the DTM calculation is independent of the best fit metallicity and hydrogen column density.
\begin{equation}
\label{eq:dtm}
\mathrm{DTM} = \frac {N\left(\mathrm{dust}\right)}{N\left(\mathrm{metals}\right)} = \frac{\sum\limits_X 10^{[X/\mathrm{H}]_{\sun}}\left(1-10^{\delta_X}\right)}{\sum\limits_X 10^{[X/\mathrm{H}]_{\sun}}} \,,
\end{equation}

As is customary in DTM analysis, we normalize our values to that of the MW. We comput a Galactic DTM using the same procedure as outlined above, assuming an $F_*$ of 0.5, as this is the average found in the 243 J09 lines of sight. We denote the MW-normalized value as \DTM. 

We calculate the error on \DTM  by propagating those from the best fit depletion through Eq. \ref{eq:dtm}. In particular, the error on the metal fraction, $10^{\delta_X}$, is $10^{\delta_X}\ln 10 ~ \alpha_{\delta_X}$, where $\alpha_{\delta_X}$ is the error on the depletion in $X$. This is then propagated in quadrature with those from the reference abundances. \\

\section{Results \label{sec:results}}

The depletion curves for each GRB-DLA are shown in Figs. \ref{fig:depletions1} and \ref{fig:depletions2}, the results of which are presented in Table \ref{table:results}, including output values for $F_*$, metallicity, and dust-to-metals ratio. 
 \subsection{Metallicity\label{subsec:metallicity}}
 
\begin{figure}
    
   \centering
   \includegraphics[width=\hsize]{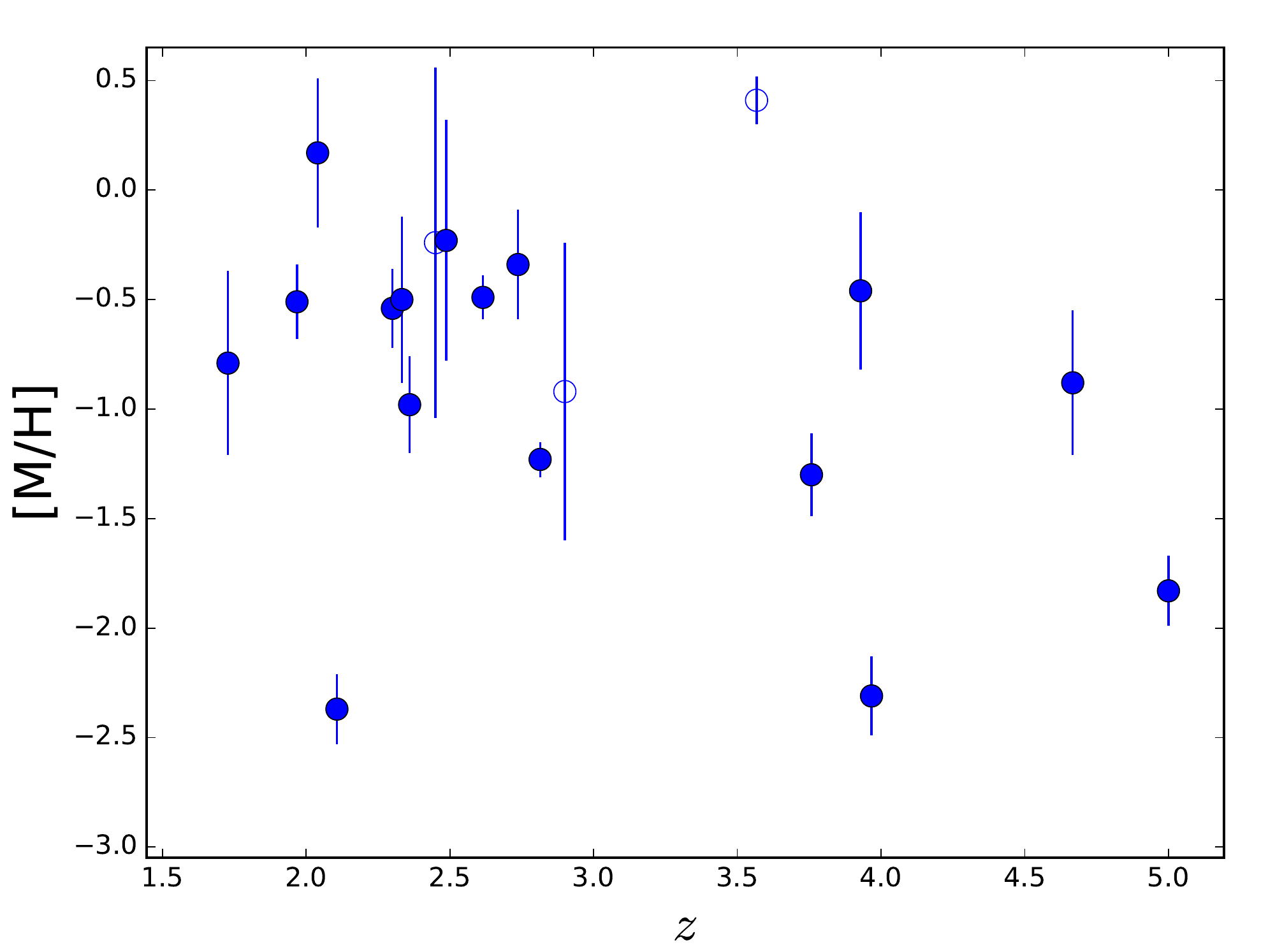}
   
   \caption{The metallicity [M/H] as a function of redshift. Open symbols are those taken from low-resolution spectra.}
              \label{fig:z_redshift}
\end{figure}

We present dust-corrected metallicities for the 19 GRB-DLAs, including 5 previously unpublished objects. The metallicities range from the very metal-poor[M/H] = -2.37  in \object{GRB 090926A} to the supersolar [M/H] = 0.41 in \object{GRB 090323}, with a median of [M/H]=-0.63, which is equal to 0.25 $Z_{\sun}$, similar to the SMC. 
The metallicities for all of the GRBs in this sample were presented by \citet{Cucchiara2015} using the apparent optical depth (AOD) method to measure column density. Our metallicities tend to agree with those from that work, although they are typically slightly higher, due to the fact that we make a correction for dust depletion. They find a weak decrease in metallicity with redshift. The metallicity as a function of redshift for our sample is shown in Fig. \ref{fig:z_redshift}. There is no significant trend, although we do note that excluding the low-resolution, supersolar data point at redshift 3.6, there could be a slight decrease in metallicity with redshift, as one would expect given the evolution of galaxies over cosmic time.

The metallicities we derive tend to have a larger uncertainty than those often quoted for GRB-DLAs. Most published metallicities do not account for dust depletion, assuming that volatile elements are good metallicity tracers as they do not deplete strongly into dust. We quote our metallicities with the knowledge that even the most volatile elements deplete to some degree, and in a linear fashion with $F_*$, thus giving rise to a larger uncertainty due to a degeneracy between [M/H] and $F_*$. We note that GRBs with detections of numerous species tend to produce a smaller uncertainty on the metallicity than those with only four, since the degeneracy can better be disentangled. 
\subsection{Dust-to-Metals ratio\label{subsec:results dtm}}
\begin{figure}
    
   \centering
   \includegraphics[width=\hsize]{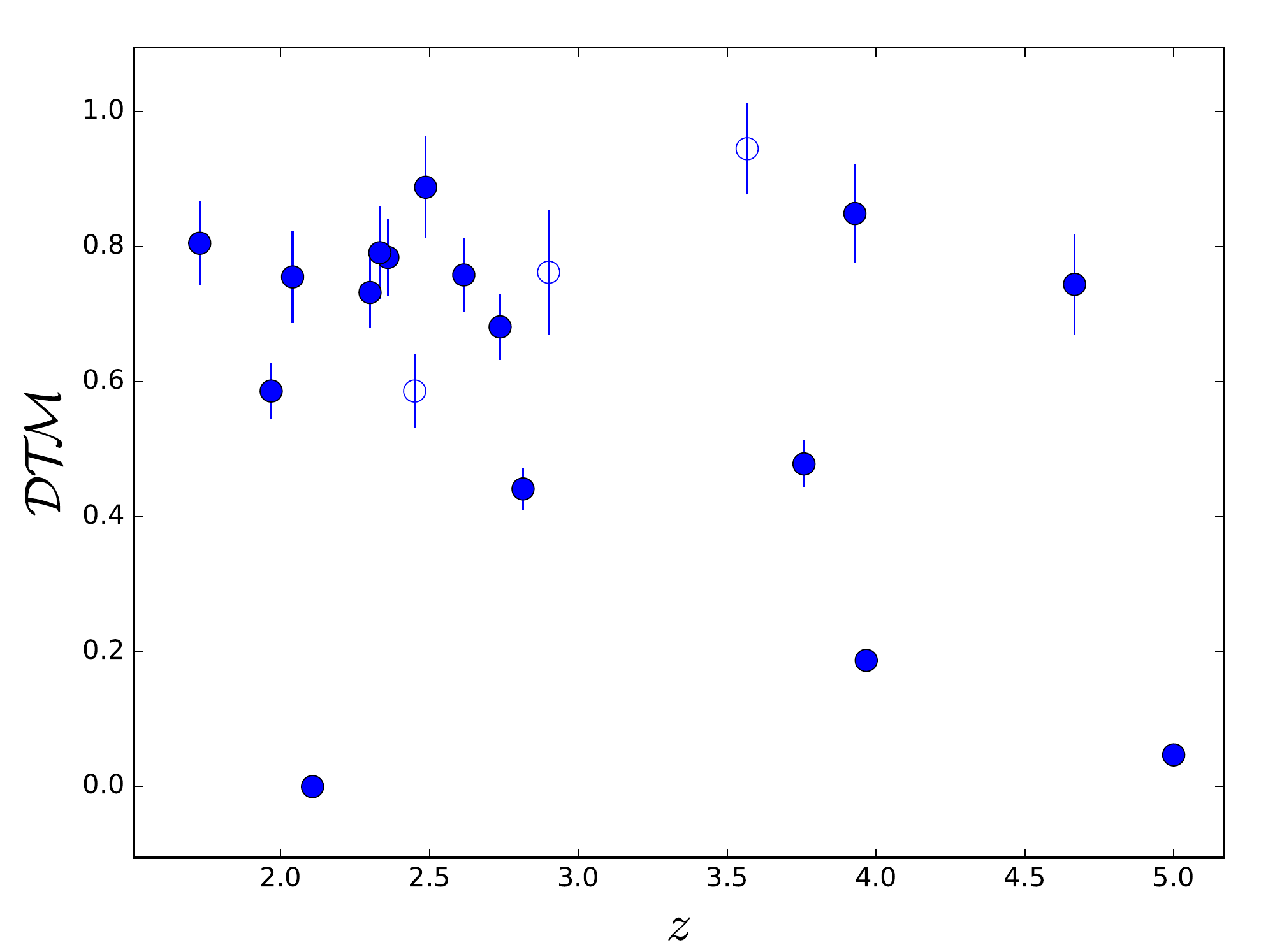}
   
   \caption{The \DTM as a function of redshift. $z$ error bars are too small to be displayed. There does not seem to be any evolution over the redshift range of 1.7-5, but no conclusions can be drawn with such a small sample.}
              \label{fig:dtm_redshift}
\end{figure}

\begin{figure}
    
   \centering
   \includegraphics[width=\hsize]{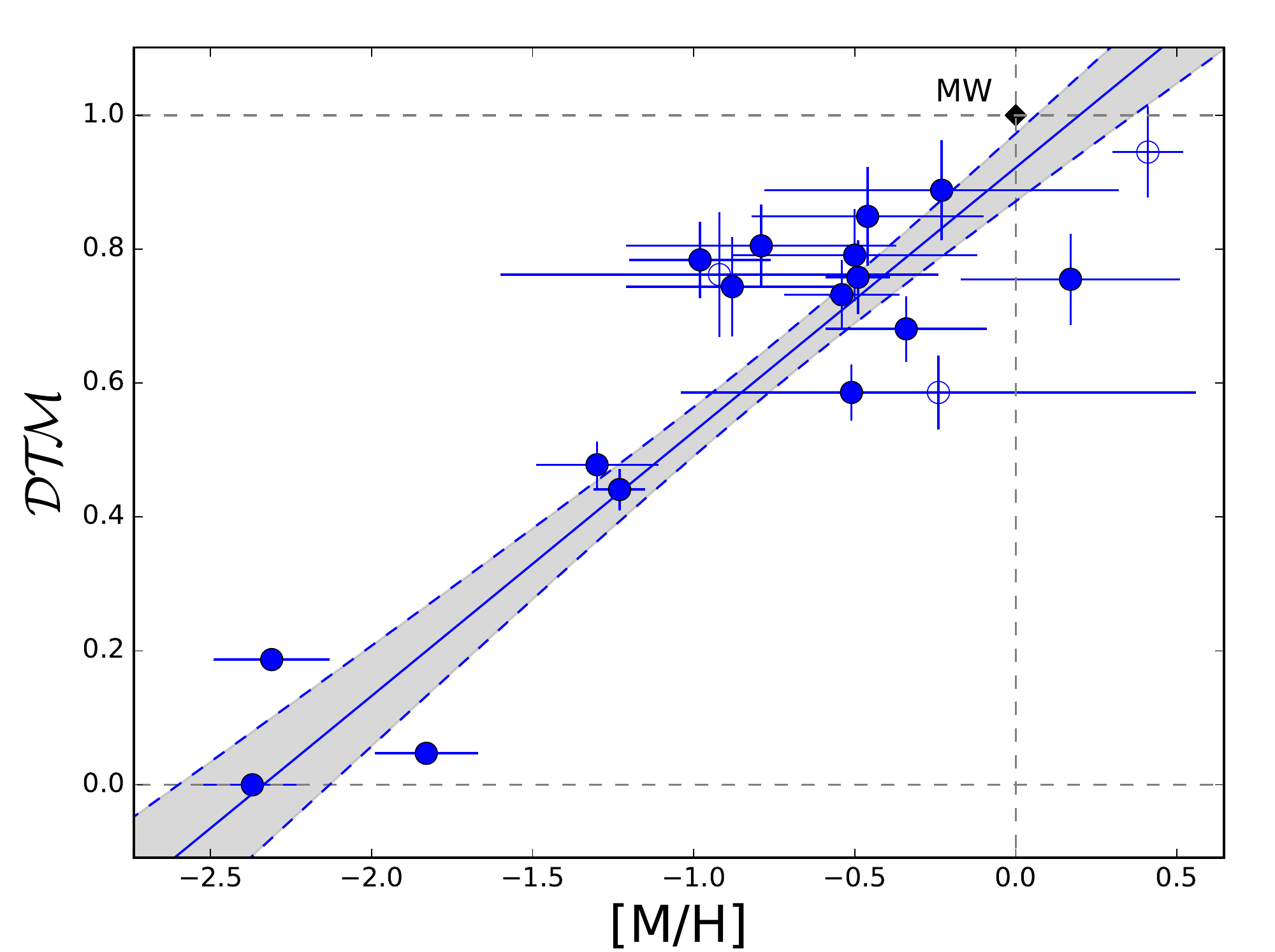} 
   
   \caption{\DTM as calculated using Eq. \ref{eq:dtm} as a function of metallicity. The dashed line and shaded area is a linear fit to the data and its 1$\sigma$ confidence interval respectively, tracing a significant positive correlation between the two variables. The trend is predominantly set by the lowest metallicity points, with those data points at [M/H]>-1 showing no obvious trend. The Milky Way is shown by a black point at [M/H]=0, \DTM=1.}
              \label{fig:dtm_Z}
    \end{figure}

 The $F_*$ values have a mean of -0.52, which equates to a [Zn/Fe] of 0.66, 0.6 dex lower than the mean Galactic value of 1.22. This is reflected in a mean \DTM of 0.62, while the median is 0.74. The standard deviation of \DTM is 0.27, such that the mean differs from the Galactic DTM by nearly $2\sigma$ while the median is lower at just below $1\sigma$ significance. This result is similar to the QSO-DLA results from \citet{DeCia2016}, whose mean and standard deviation are 0.70 and 0.26 respectively. 
\object{GRB 090926A} is the only DLA which shows no dust depletion with $F_* = -1.7$, which lies to the left of the point where all the depletion slopes cross the axis of zero depletion in Fig. \ref{fig:f*}.\\
\\
 Fig. \ref{fig:dtm_redshift} shows the \DTM plotted against redshift. From this small sample, we don't see any significant trend with redshift. This result is consistent with the hydrodynamical simulations of \citet{McKinnon2016}, who find no evolution in the DTM at redshifts $z \geq 1$.
In Fig. \ref{fig:dtm_Z}, we plot our \DTM against metallicity, and find a positive correlation between the two with a Spearman's Rank of $\rho=0.63$, which with 19 data pairs leads to a false-correlation probability of 0.004. There is a potential flattening of the relation, such that above [M/H] $= -1$ there is no real correlation. The potential reasons for this are discussed in the following section.

 \section{Discussion\label{sec:discussion}}

\subsection{The Origin of Dust\label{subsec:origin}}

Fig. \ref{fig:dtm_Z} shows that the dust-to-metals ratio increases with metallicity. Typically, this would support the view that the dust is formed by grain growth in the ISM \citep{Draine2009}. \citet{Mattsson2014} can also explain this scenario in terms of dust created in stars and supernovae, and then kept in balance by ISM grain growth counteracting dust destruction mechanisms. We also notice a potential flattening of this trend above metallicities of 0.1$Z_{\sun}$. This flattening is the opposite to what is predicted in the models of \citet{Mattsson2014} and \citet{Mattsson2016} where the DTM is fairly constant (and low) until a critical metallicity of 0.1 $Z_{\sun}$ at which point dust production via grain growth is kick-started and the DTM grows towards the Galactic value at solar metallicity.
Our observed trend can instead be explained by a higher rate of dust destruction at lower metallicities. As noted in \citet{DeCia2013}, this could be due to the  star forming environments that GRBs are known to trace (e.g. \citealt{Savaglio2009}), which are somewhat different to the solar-like environment of \citet{Mattsson2016}. The strong radiation fields in such environments produced by young OB stars and shocks from supernovae (SNe) may destroy dust faster than it can be created by grain growth when the metallicity is low.
Another possibility is that the GRB-DLAs are actually tracing dust in the CGM rather than the ISM of their host galaxies where dust evolution is likely to be somewhat different to that in the ISM models referred to previously. Typically, however, the DLAs are located a few hundred parsecs from the GRB site itself (e.g. \citealt{Hartoog2013,DElia2014,Friis2015}), and thus located in the ISM of the GRB's host galaxy.

The amount of dust along the line of sight can also be measured by the effect it has on the spectral energy distribution (SED) of a GRB afterglow. Specifically, the SED is `reddened', and this reddening can be expressed as the total extinction in the $V$- band, $A_V$.
The values for $A_{V;\mathrm{SED}}$ are found by fitting broadband SEDs from the optical to the X-ray regimes (see e.g. \citealt{Greiner2011,Schady2012,Covino2013}; and upcoming papers Bolmer et al. and Greiner et al, both \textit{in prep}). For all bursts from 2007 onwards, we use optical/NIR data from the seven-channel imager GROND \citep{Greiner2008}, and X-ray data from the X-ray Telescope (XRT; \citealt{Burrows2005}) on board \textit{Swift} \citep{Gehrels2004}. For the pre-GROND bursts, see \citet{Schady2011}. A simple power-law or broken power-law is fit to the observed data, and `missing' flux in the bluer visible bands is attributed to dust. This reddening is fit with one of three different exctintion laws, namely those from the SMC, LMC, and MW \citep{Pei1992}, and is described by the colour excess, $E(B-V)$. This is converted into $A_{V;\mathrm{SED}}$ via the relation 
\begin{equation}
A_V = R_VE(B-V)\,,
\label{eq:av}
\end{equation}
where $R_V$ is the total-to-selective extinction, and is fairly well known for the Local Group extinction curves at an average of 3.08, 3.16, and 2.93 for the MW, LMC and SMC, respectively. It includes silicates and carbonaceous grains and depends largely on the grain-size distribution. Typically, these Local Group extinction laws produce a good fit to GRB SEDs \citep{Schady2010,Kann2010,Greiner2011},  although a more complex dust model might fit the extinction curves better. Although the best fit $A_V$ varies slightly depending on which curve is used, the use of NIR and X-ray data in the SED fit typically provides good constraints on $A_{V;\mathrm{SED}}$. The extinction law used for the final $A_{V;\mathrm{SED}}$ measurement is that which results in the best $\chi^2_{\mathrm{red}}$ value.

 In Fig. \ref{fig:zafar_dtm}, we plot the metals-to-dust ratio according to the definition of \citet{Zafar2013}, which uses the $A_V$ as a dust tracer. As in  that work, we see no strong trend with metallicity, at odds with the result from Fig. \ref{fig:dtm_Z}. We  note than our mean metals-to-dust of 21.65 cm$^{-2} A_V$mag$^{-1}$ is higher than that from their sample, and we see a higher spread of $\sigma=0.46$ dex. A Spearman's rank test gives $\rho = 0.38$ with a false positive probability of $P=0.11$, suggesting that there is perhaps a slight positive correlation, and indeed in the opposite direction to that in our \DTM method.  Given that the metal measurement  comes from the same place in both methods, there must be a discrepancy between how the dust is measured, the reasons for which we explore in the following sections. 

\subsection{The $A_{V;\mathrm{SED}}$ to $A_{V;\mathrm{DTM}}$ discrepancy\label{subsec:compare av}}

\begin{figure}
    
   \centering
 
 \includegraphics[width=\hsize]{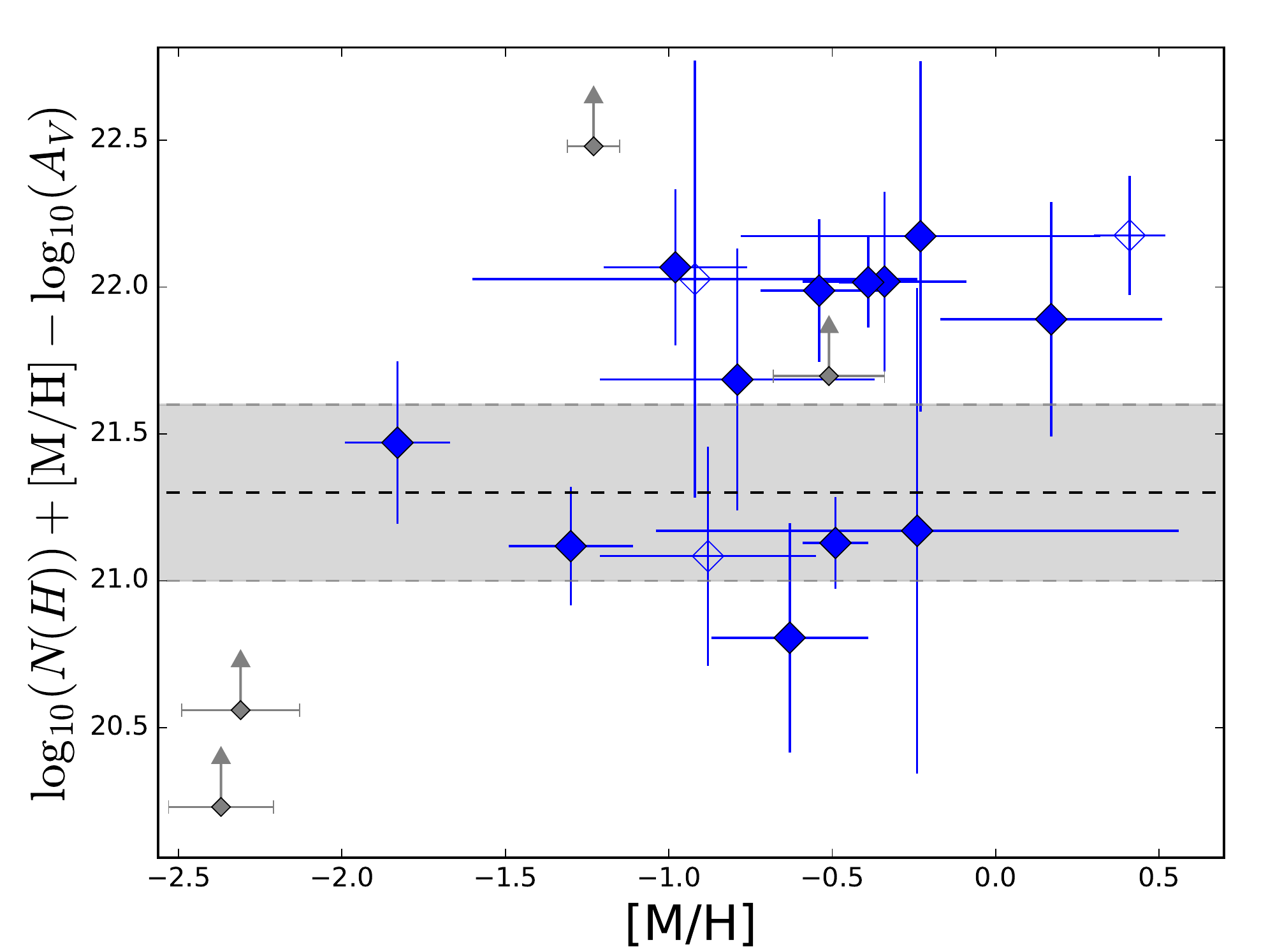} 
   
   \caption{Metals-to-dust diagnostic used by \citet{Zafar2013}, as applied to our data along with the Local Group average and standard deviation that they quote (dashed line and shaded area). }
              \label{fig:zafar_dtm}
    \end{figure}

We can see from Figs. \ref{fig:dtm_Z} and \ref{fig:zafar_dtm} that depletion and extinction seem either to have different sensitivity, or  not to trace the same dust along the line of sight, or properties thereof. To compare these values we look to the relation used to calculate a value of $A_V$ from a depletion-measured DTM, which we label $A_{V;\mathrm{DTM}}$. This is based on the average extinction for a given hydrogen column density in the MW, scaled for DTM and metallicity, as per \citet{Savaglio2003}, and using the $N(\mathrm{H})/A_V$ from \citet{Watson2011} 
\begin{equation}
\label{eq: av from dtm}
 A_V = 0.45\frac{DTM}{DTM_{\mathrm{Gal}}}\frac{Z}{Z_{\sun}}\frac{N\left(\mathrm{H}\right) \mathrm{cm}^{-2}}{10^{21}}~\mathrm{mag}\,,
\end{equation}
with $N$(H) measured in $\mathrm{cm}^{-2}$.
In the literature there are many cases of GRB afterglows where a direct $A_V$ measurement from the SED was possible, as well as spectra with measurable depletion, and there is often disagreement between the two values, with the depletion-inferred $A_V$ usually higher than the SED value (e.g. \citealt{Watson2006,Savaglio2012,Friis2015}). For our sample, we compare our independently measured $A_{V;\mathrm{SED}}$ values to $A_{V;\mathrm{DTM}}$ based upon the \DTM, $N$(H) and [M/H] from our fits, the result of which is shown in Fig \ref{fig:both_avs_notlog}. There seem to be two distinct categories of objects: group (1) are found above the green 1:1 line and make up the majority of the sample and  show the known overprediction of $A_{V;\mathrm{DTM}}$ compared to $A_{V;\mathrm{SED}}$, which are best fit by the blue dashed line;   group (2) are found below this line and are those whose $A_{V;\mathrm{DTM}}$ prediction is lower than that measured from the SED. These objects include the known outlier \object{GRB 070802} \citep{Kruehler2008,Eliasdottir2009} at an $A_{V;\mathrm{SED}}$ of 1.23 mag. This  underprediction for \object{GRB 070802} is also noted by \citet{DeCia2016}, and could be a result of the uncertain column density measurements resulting from low-resolution spectral data.

Including those GRBs with $A_{V;\mathrm{SED}}$ upper limits, 11   are categorized as  overpredictions in group (1), while 6 are definitely in group (2). Of these, one is \object{GRB 070802}. The others are GRBs \object{050820A}, \object{100219A}, \object{111008A}, \object{130408A,} and \object{141028A}. \object{GRB 090926A} shows negligible dust in both depletion and extinction. 
 
\subsection{Accuracy of $A_{V;\mathrm{DTM}}$ - depletion as a reliable tracer of dust}

A potential reason for the discrepancy is touched upon in \citet{Zafar2013}, who mention that while depletion is often based upon Fe, the vast majority of dust mass is composed of O, C, Mg, and Si, arguing that Fe-based depletion measurements may not correctly trace most of the dust. Indeed, \citet{Dwek2016} provide a compelling argument for Fe dust production being different to that of the bulk of the elements; specifically,  it is   formed by cold grain growth in the ISM rather than in CCSNe and post-AGB star envelopes. In this argument, depletion based upon iron measurements is therefore bound to trace different dust to extinction. However, the linear depletion sequences seen by \citet{Jenkins2009} and \citet{DeCia2016} show that by calculating the $F_*$ for a particular line of sight, we can get a robust prediction for dust fractions and columns for all elements, including those not measured in the DLA, and thus we  incorporate not only the dust compounds traced by iron, but  the silicates and carbonaceous grains as well. 

We have followed the procedure in \citet{DeCia2016} (see Sect. \ref{subsec:ref_abundances}) to correct relative abundances for nucleosynthetic effects, such that any discrepancies between our adopted intrinsic abundance and that true to the DLA are likely to be marginal, and certainly not large enough to cause the observed offset in the  $A_{V;\mathrm{DTM}}$ prediction.

\subsection{Accuracy of $A_{V;\mathrm{SED}}$ \label{subsubsec: av robust}}

Assuming that our depletion measure is indeed a solid representation of the total dust column, we look to $A_{V;\mathrm{SED}}$ for the reason  why there could be a discrepancy. The question  is whether the Local Group extinction curves are a good fit for GRB-DLAs or whether  something other than the MW, LMC, or SMC should be used as their model,  such as grey dust \citep{Perley2008}. Indeed, \citet{Friis2015a} claim that grey dust extinction, so called because the extinction is weakly dependent on wavelength, could be prominent in up to  25\% of GRB-DLAs, including \object{GRB 121024A} which is included in our sample. Their reasoning is that a top-heavy grain-size distribution would cause a very flat extinction curve. When fitting a broken power-law SED, there is then a degeneracy between the steepness of the extinction curve (i.e. the $A_V$), and the position of the break between X-ray optical power-law slopes \citep{Schady2012};  one of the solutions is a large amount of grey dust, which corresponds to a large $A_V$ for a small $E$(B-V).
 If the extinction in our group (1) DLAs is   caused by grey dust, then we may be able to reconcile the overpredicted $A_{V;\mathrm{DTM}}$ with $A_{V;\mathrm{SED}}$.
However, some of these DLAs show compelling evidence for the contrary: \\
\object{GRB 120327A} is best fit by a simple power law and an SMC-like extinction of $A_{V;\mathrm{SED}} = 0.05$ mag, with $A_{V;\mathrm{DTM}}= 0.18$ mag. The power law removes any degeneracy in the slope of the dust extinction law, and thus excludes the possibility of significant grey dust. Another example is \object{GRB 120815A}, whose SED is fit by a power law and SMC-like extinction to give an $A_{V;\mathrm{SED}}$ value of 0.08 mag, which is significantly smaller than the $A_{V;\mathrm{DTM}}$ of 0.44 mag. We find that \object{GRB 121024A} is also best fit with a power law. Indeed, Fig. \ref{fig:both_avs_notlog} shows that only one object that has an overpredicted $A_V$ was fit with a broken power law.

Although there is strong evidence for dust destruction caused by the GRB itself \citep{Morgan2014}, this would not cause a discrepancy between $A_{V;\mathrm{SED}}$ and $A_{V;\mathrm{DTM}}$. Any dust that extinguishes the GRB would also be visible in depletion, so if that dust is destroyed it is no longer visible in  depletion or extinction.

Intervening systems such as $\ion{Mg}{II}$ absorbers are known to contain similar quantities of dust as galaxies (e.g. \citealt{Menard2012}), although QSO-DLAs tend to show very little reddening \citep{Krogager2016,DeCia2016}. For the group (2) objects, we notice that \object{050820A}, \object{100219A}, \object{111008A}, and \object{130408A} all have intervening absorbing systems. Should these objects have a high dust content, they could significantly affect the SED of the GRB afterglow, such that the reddening caused by dust in the host galaxy itself is indeed smaller, and thus pushes these objects towards the 1:1 line. 
However, for these intervening systems to be the reason for a much higher $A_{V;\mathrm{SED}}$ than from DTM, they would need to contribute around 80\% of the extinction along the line of sight, whereas the systems in our sample are much weaker in metal line absorption than the host DLA (e.g. in GRB 100219A, \citealt{Thoene2013}). We therefore find it unlikely that a significant amount of the extra extinction is caused by intervening systems.

\subsection{ Equivalent dust column density \label{subsubsec: av_dtm}}

\begin{figure}
    
   \centering
   \includegraphics[width=\hsize]{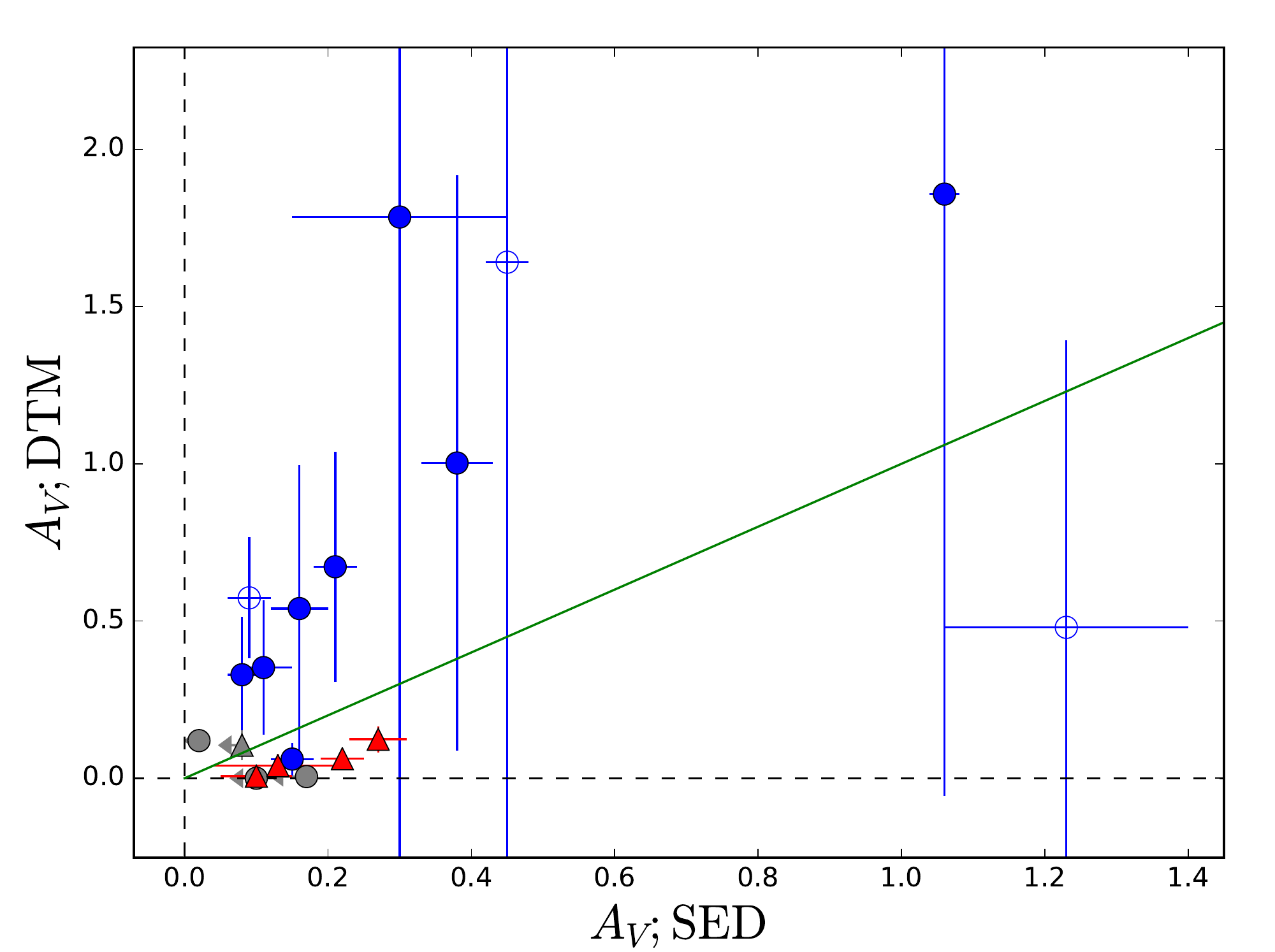} 
   
   \caption{$A_V$ as measured directly from the SED against that calculated from the DTM. The green line indicates a 1:1 conversion between the two. Blue circles represent SEDs best fit with a simple power-law, whereas red triangles are used for broken power-law fits. Empty points refer to depletion measured from low-resolution spectra. Large error bars are due to significant errors in several measured parameters being combined in quadrature.}
              \label{fig:both_avs_notlog}
    \end{figure}

Having established that depletion is a good tracer of the dust, and with $A_{V;\mathrm{SED}}$ being accurate and reliable, we look to the relation used to calculate $A_{V;\mathrm{DTM}}$, Eq. \ref{eq: av from dtm}. This is based upon the relation between hydrogen and dust in the Galaxy, where a column of $N(\mathrm{H})=10^{21}$ cm$^{-2}$ results in an $A_V$ of 0.45. We note that the value of the Galactic gas-to-dust ratio varies depending on the sample and technique used to measure it. We use the result of \citet{Watson2011}. Measurements of this value have been consistent over the past few decades, and include those by \citet{Bohlin1978}, \citet{Predehl1995}, and \citet{Guever2009}. The value used does not alter the fact that a significant discrepancy is observed. In DLAs, the hydrogen column density is scaled for dust-to-metals ratio and metallicity to take into account the differing dust-to-gas ratios in such environments. However, the discrepancies in Fig. \ref{fig:both_avs_notlog} show that the scaling between this equivalent dust column density and the $A_V$ may well be incorrect. That is to say that in DLAs, such a column of dust does not have as much of a reddening effect as in the MW (see e.g. \citealt{Campana2009}.) This would indeed be solved by the make-up of the dust being different, but this is hard to explain given the well-determined extinction laws that are observed in GRB afterglows, which are consistent with the Local Group extinction laws (\citealt{Schady2011,Greiner2011,Covino2013}). 
One could argue that the problem arises from using a MW scaling relation with $A_{V;\mathrm{SED}}$ measurements based upon mostly SMC-like extinction laws, but both Magellanic Clouds have similar a $A_{V}/N_{\mathrm{H}}$ to that in the MW \citep{Zafar2013,Watson2011}.

 The reason for the scatter thus remains unclear, and we therefore advise significant caution against basing $A_V$ predictions on the $A_V$-to-$N$(H) ratio of the MW and Local Group. We also suggest that the discrepancy between the different methods of quantifying dust is the reason for the disagreement between the trend, or non-trend, seen in DTM with metallicity.

\section{Conclusion\label{sec:conc}}

Gamma-ray bursts are a unique if  somewhat biased probe of the dust-to-metals ratio in the high-redshift Universe. 
GRBs occur only within certain types of galaxies \citep{Kruehler2015,Perley2016a,Perley2016d}, and thus are not totally unbiased probes, although this effect is reduced as redshifts greater than around 2 are reached \citep{Perley2013,Greiner2015a,Schulze2015}. They are also complementary to QSO-DLAs, and this work expands our observational knowledge of the DTM into the inner regions of galaxies in the distant Universe. We have used optical/NIR spectroscopy from a sample of 19 GRB afterglows in order to measure the metal and dust content of the DLAs in their host galaxies, including previously un-published metal column densities and metallicities for five objects. By using dust depletion models based on the MW, as well as QSO-DLAs, we have used a thorough method to determine the column densities of dust and of metals in order to calculate a dust-to-metals ratio. We find that the DTM follows a positive trend with metallicity, supporting the theory that a significant amount of dust is formed in situ in the ISM. 
We have investigated the discrepancy between the results of \citet{DeCia2013} and \citet{Zafar2013}, concluding that $A_{V;\mathrm{SED}}$ and depletion are not analogous measurements of dust. We see the common trend that $A_{V;\mathrm{DTM}}$ is often higher than $A_{V;\mathrm{SED}}$, which we tentatively suggest could be due to the scaling between depletion-measured DTM and $A_V$ being different in GRB host galaxies to the MW. We also note a significant number of objects whose $A_{V;\mathrm{DTM}}$ values are underpredictions compared to $A_{V;\mathrm{SED}}$, and despite seeing what looks like two distinct populations, we are unable to satisfactorily reconcile the two using theories such as grey dust or intervening systems. We thus suggest that, given the large scatter between the two, DTM measured from depletion should not be used as a proxy for $A_V$, and encourage further work with larger samples to investigate the problem further.

\begin{acknowledgements}
We thank the anonymous referee for the thorough feedback and detailed comments which have significantly enhanced the strength of this paper. We are grateful for the support of the GROND GRB team. We thank Annalisa De Cia and Darach Watson for interesting and useful discussions. We thank C. Ledoux and P. Vreeswijk for providing a compilation of oscillator strengths. P.W, P.S, and R.M.Y acknowledge support through the Sofja Kovalevskaja Award to P.S from the Alexander von Humboldt Foundation of Germany. Part of the funding for GROND (both hardware and personnel) was generously granted from the Leibniz-Prize to Prof. G. Hasinger (DFG grant HA 1850/28-1).
\end{acknowledgements}

\bibliographystyle{aa} 
\bibliography{all_papers} 

\renewcommand\thesection{\Alph{section}}
\onecolumn
\newpage
\begin{appendix}
\section{}
\setcounter{figure}{0} \renewcommand{\thefigure}{A.\arabic{figure}}
\begin{figure}[h]
\centering
\includegraphics[width=18cm]{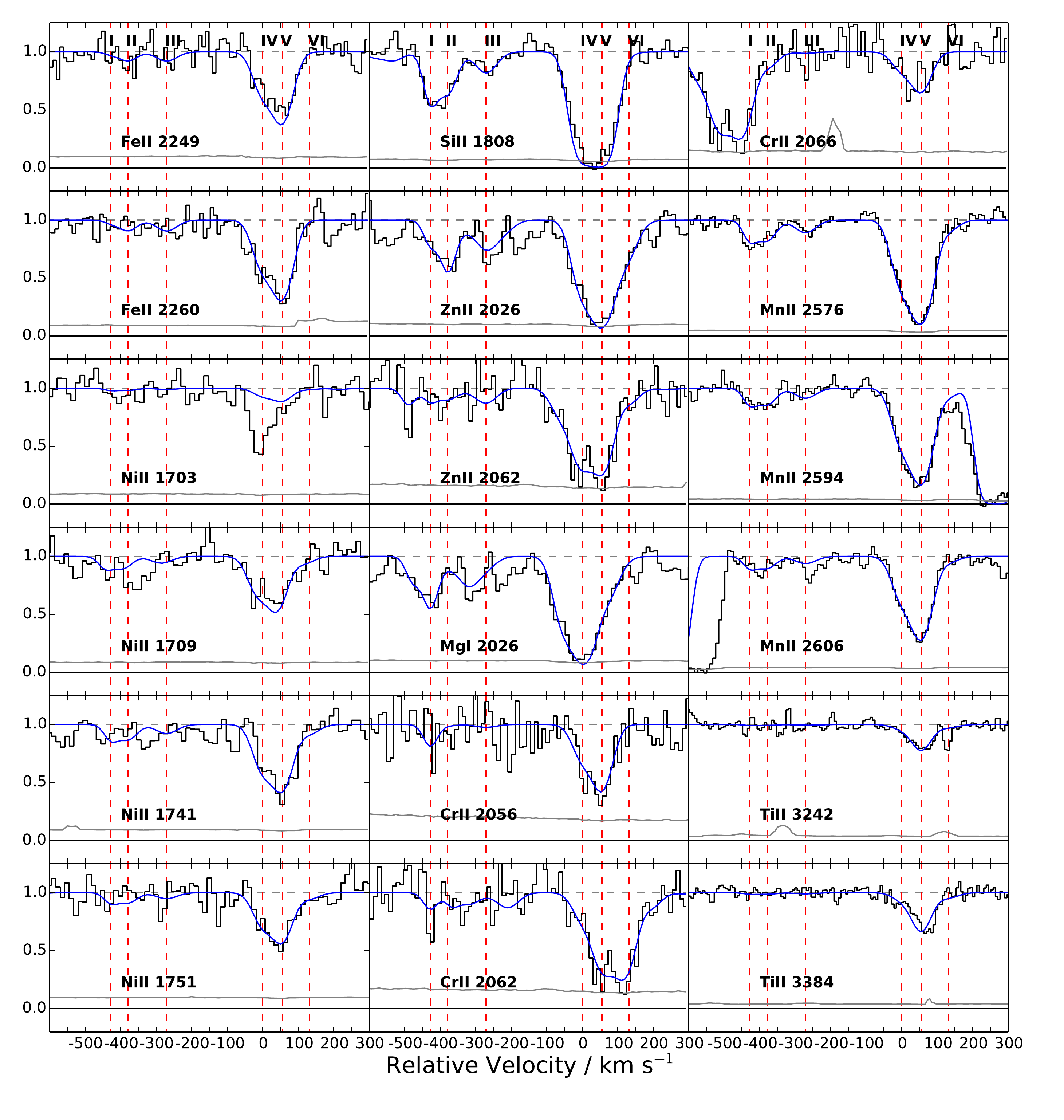}
\caption{Metal absorption lines in the X-Shooter spectrum of GRB 120119A.  Red dashed lines indicate velocity components. Resolution in VIS arm: $v = 31.4 $km s$^{-1}$. }
\label{lines120119}
\end{figure}

\newpage
\newpage
\begin{figure}[h]
\centering
\includegraphics[width=18cm]{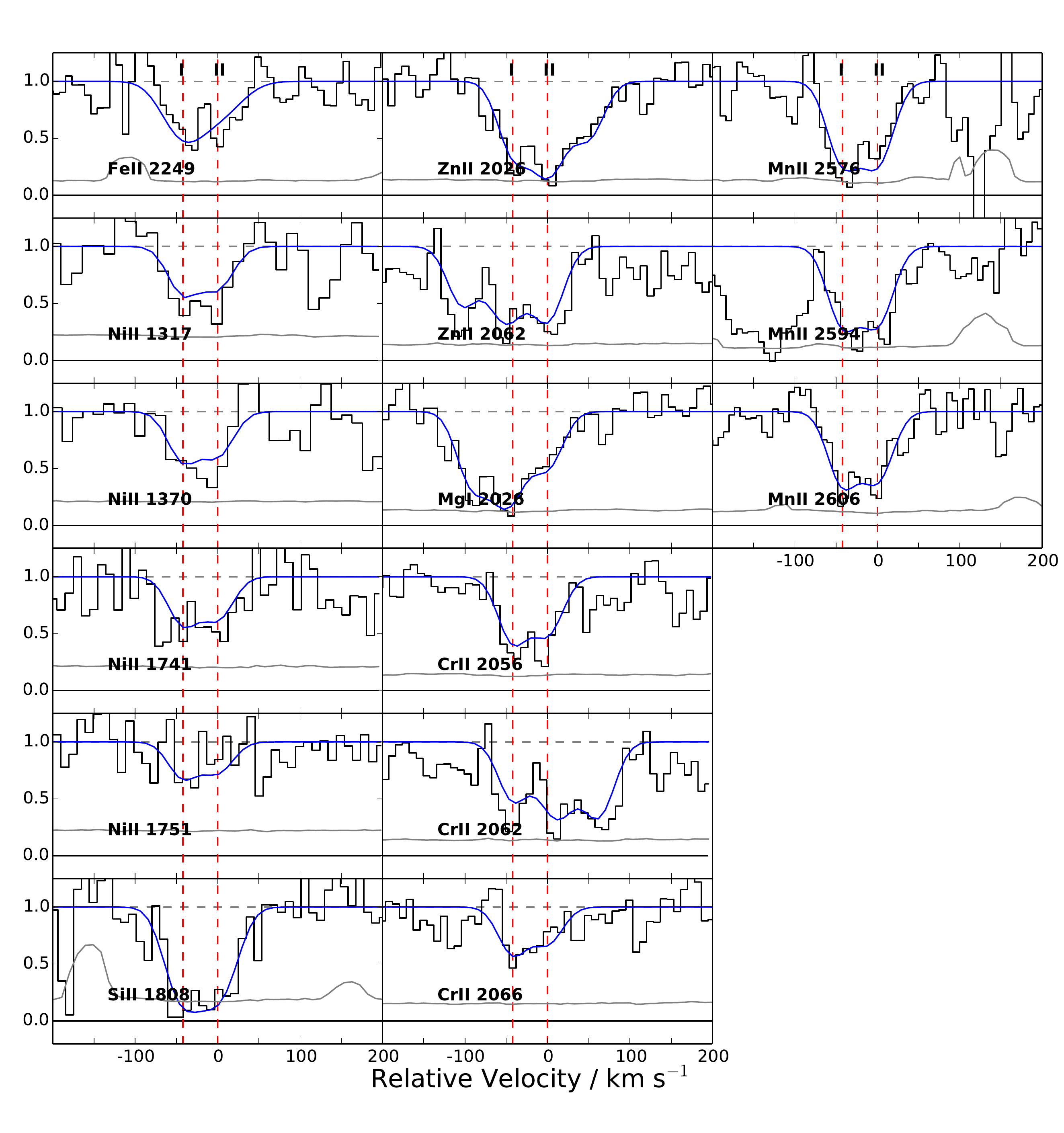}
\caption{Metal absorption lines in the X-Shooter spectrum of GRB 120716A.  Red dashed lines indicate velocity components. Resolution in VIS arm: $v = 35.0 $km s$^{-1}$. }
\label{lines120716}
\end{figure}

\newpage
\newpage
\begin{figure}[h]
\centering
\includegraphics[width=18cm]{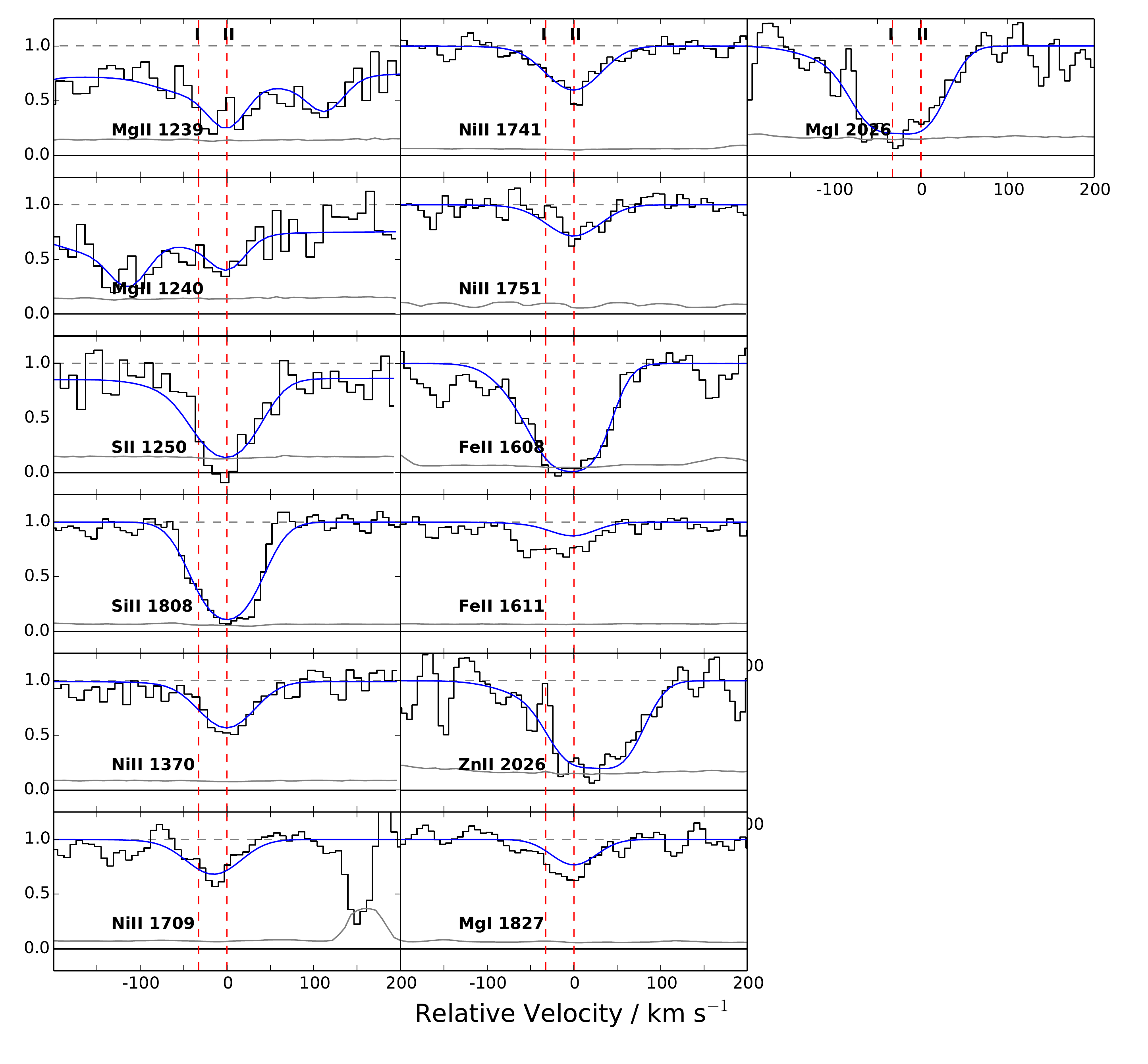}
\caption{Metal absorption lines in the X-Shooter spectrum of GRB 120909A.  Red dashed lines indicate velocity components. Resolution in VIS arm: $v = 30.1 $km s$^{-1}$. 
The continuum in $\ion{Mg}{II} 1239/1240$ and $\ion{S}{II}$ 1250 is affected by the Lyman-$\alpha$ red damping wing. The extra absorption apparent in \ion{Fe}{II} 1611 is from the excited state transition \ion{Fe}{II} 5s 1612. }
\label{lines120909}
\end{figure}

\newpage
\newpage
\begin{figure}[h]
\centering
\includegraphics[width=18cm]{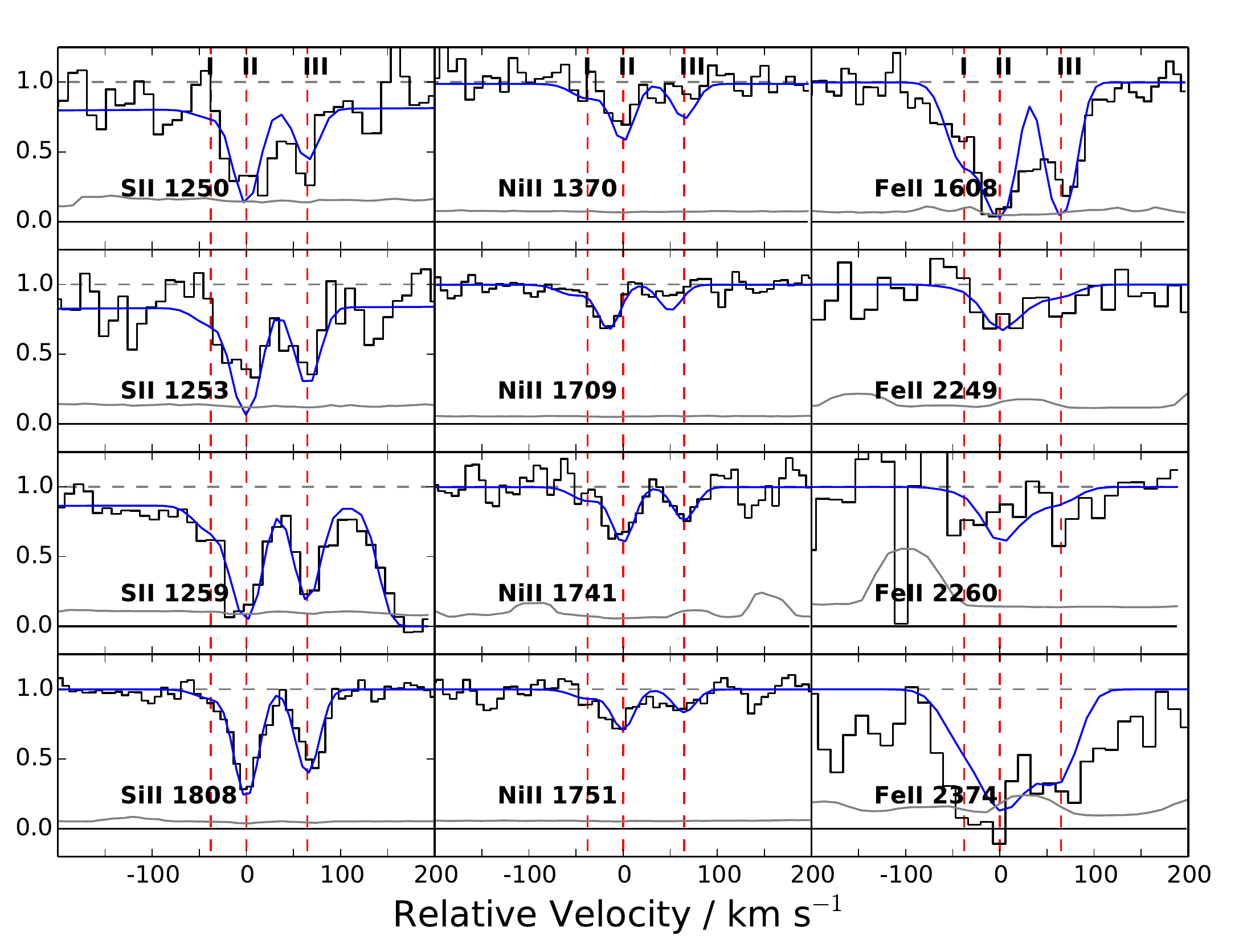}
\caption{Metal absorption lines in the X-Shooter spectrum of GRB 130408A. Red dashed lines indicate velocity components. Resolution in VIS arm: $v = 20.0 $km s$^{-1}$. The continuum in $\ion{S}{II}$ 1250/1253/1259 is affected by the Lyman-$\alpha$ red damping wing.}
\label{lines130408}
\end{figure}

\newpage
\newpage
\begin{figure}[h]
\centering
\includegraphics[width=18cm]{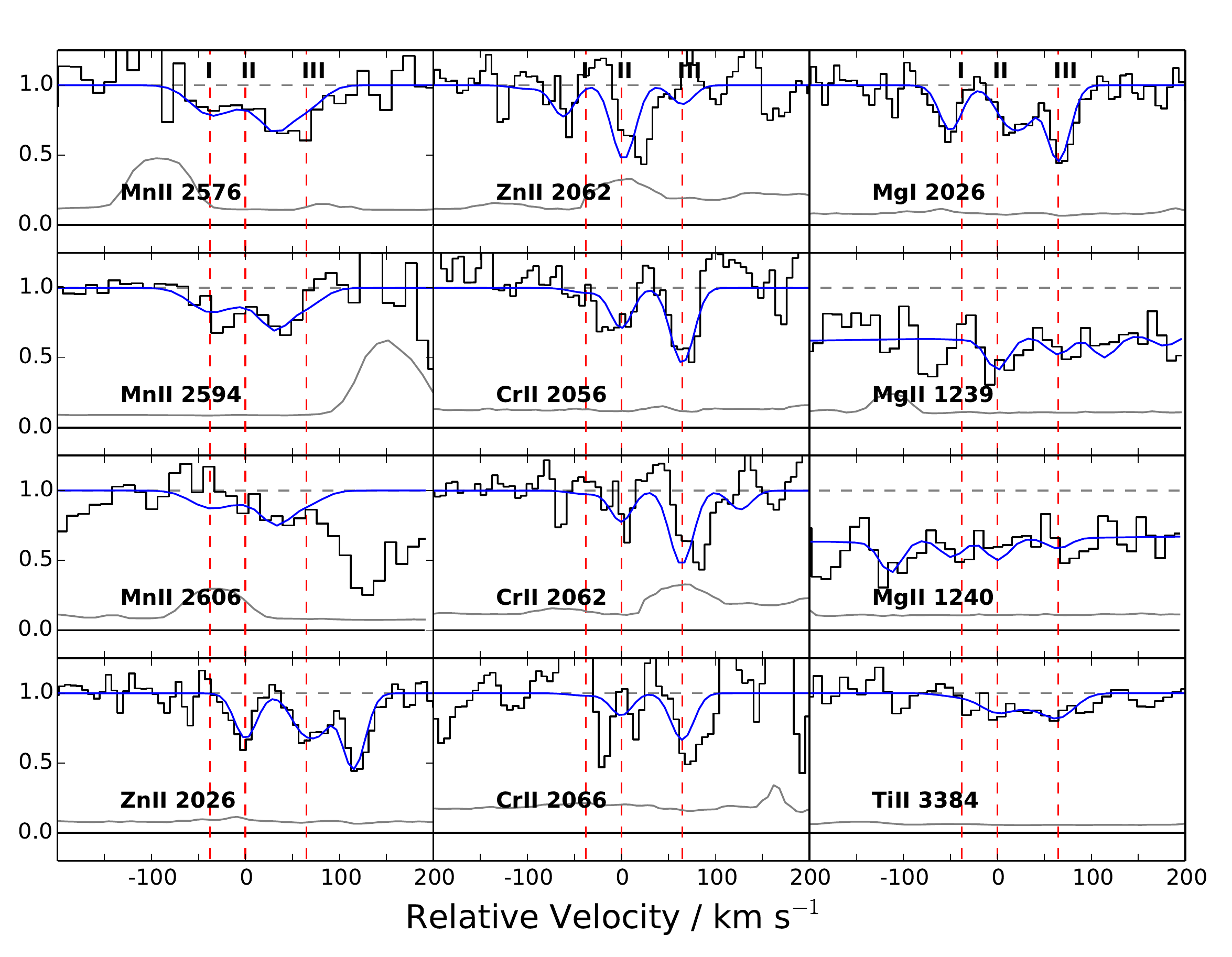}
\caption{Metal absorption lines in the X-Shooter spectrum of GRB 130408A. Red dashed lines indicate velocity components. Resolution in VIS arm: $v = 20.0 $km s$^{-1}$. The absorption to the right of the $\ion{Mn}{II}$ 2606 line is due to telluric absorption.}
\label{lines130408b}
\end{figure}

\newpage
\newpage
\begin{figure}[h]
\centering
\includegraphics[width=18cm]{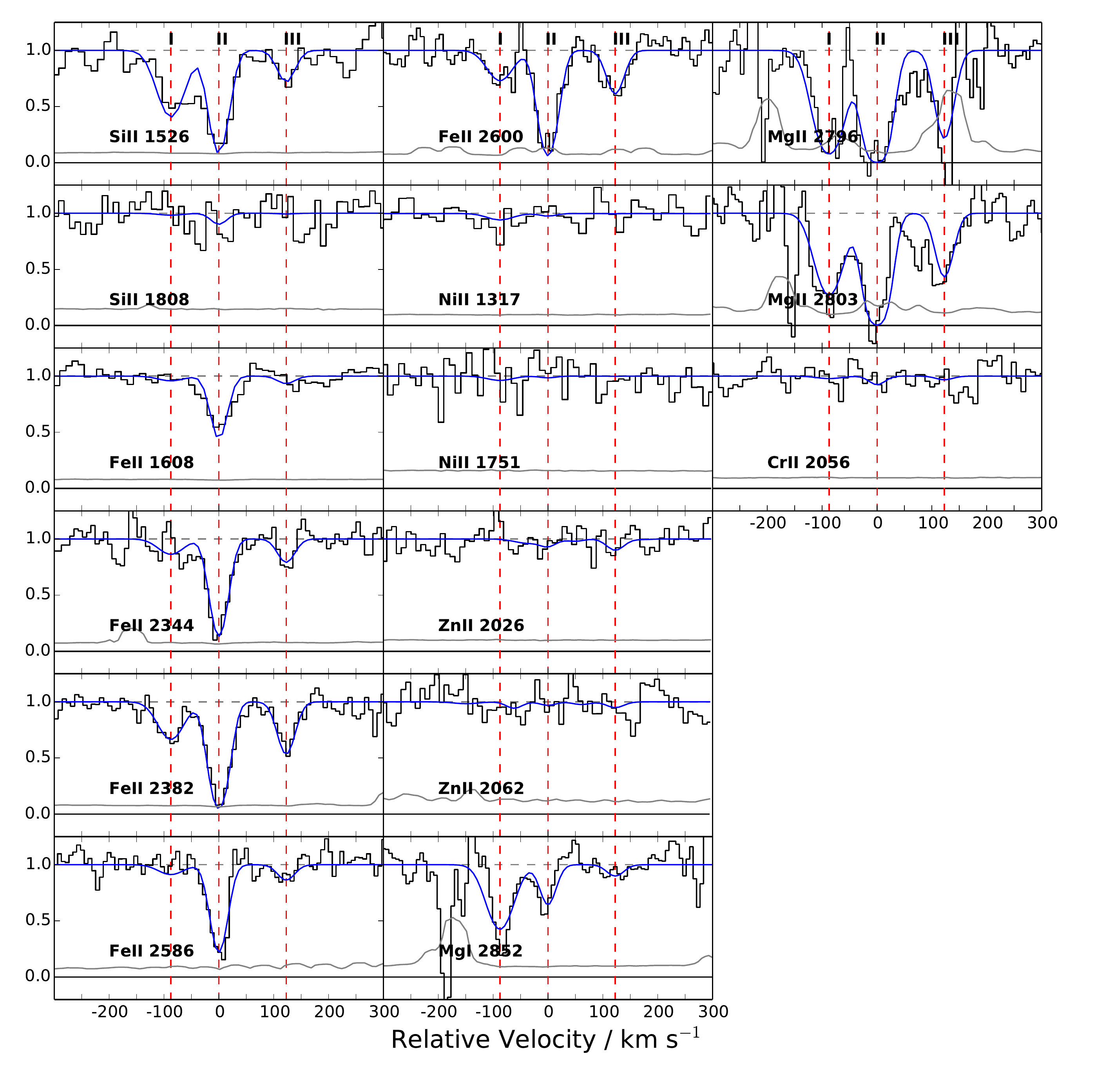}
\caption{Metal absorption lines in the X-Shooter spectrum of GRB 141028A.  Red dashed lines indicate velocity components. Resolution in VIS arm: $v = 25.0 $km s$^{-1}$. The narrow saturated lines in the $\ion{Mg}{II}$ 2796/2803 and $\ion{Mg}{I}$ 2852 are due to telluric absorption.}
\label{lines141028}
\end{figure}
\newpage
\section{}                              
                                                                                                
\setcounter{figure}{0} \renewcommand{\thefigure}{B.\arabic{figure}}
\twocolumn
\begin{figure}[h]
\centering
\includegraphics[width=9cm]{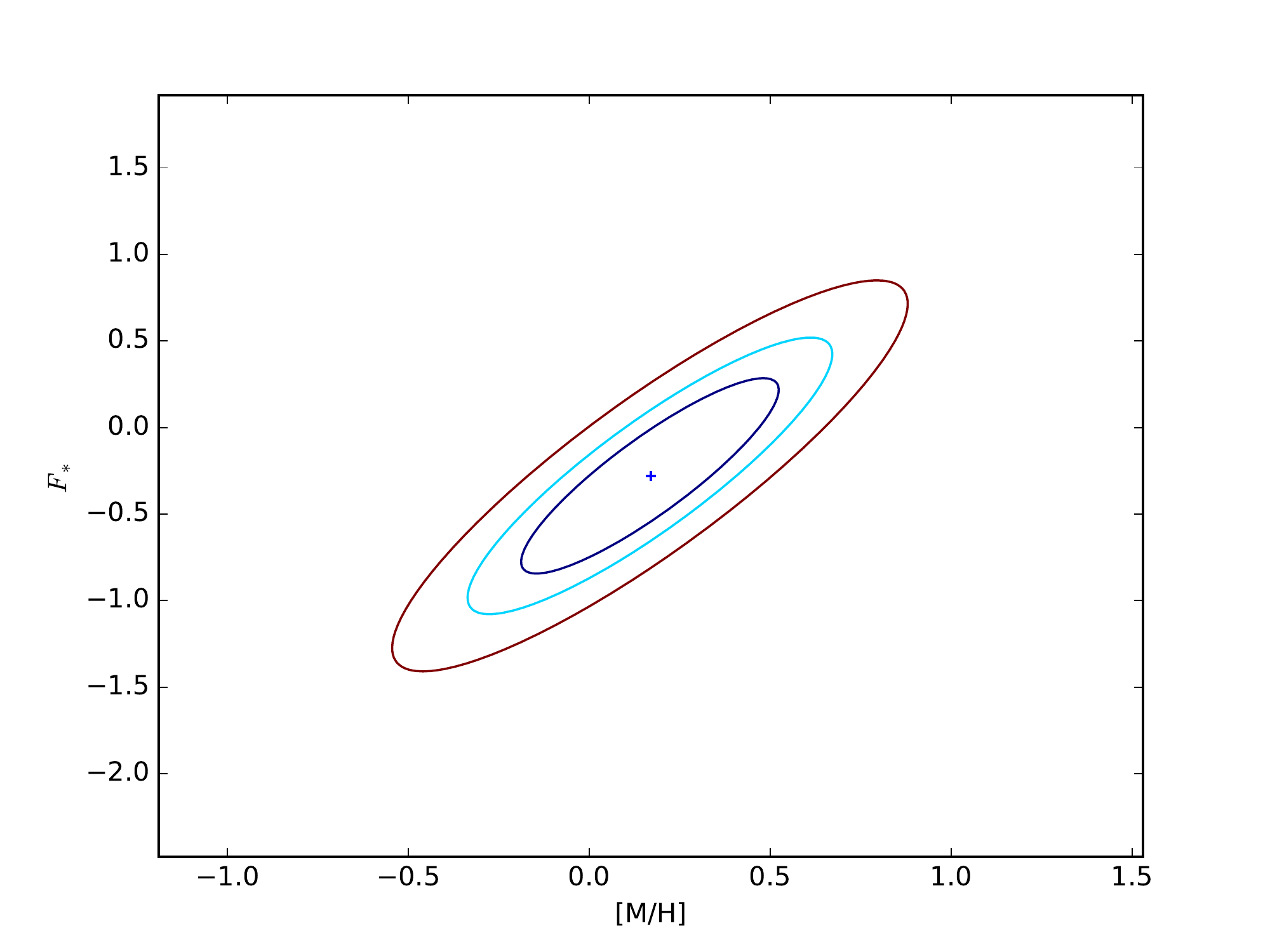}
\caption{Confidence regions for $F_*$ and [M/H] for GRB 000926.}
\label{contours000926}
\end{figure}
\begin{figure}[h]
\centering
\includegraphics[width=9cm]{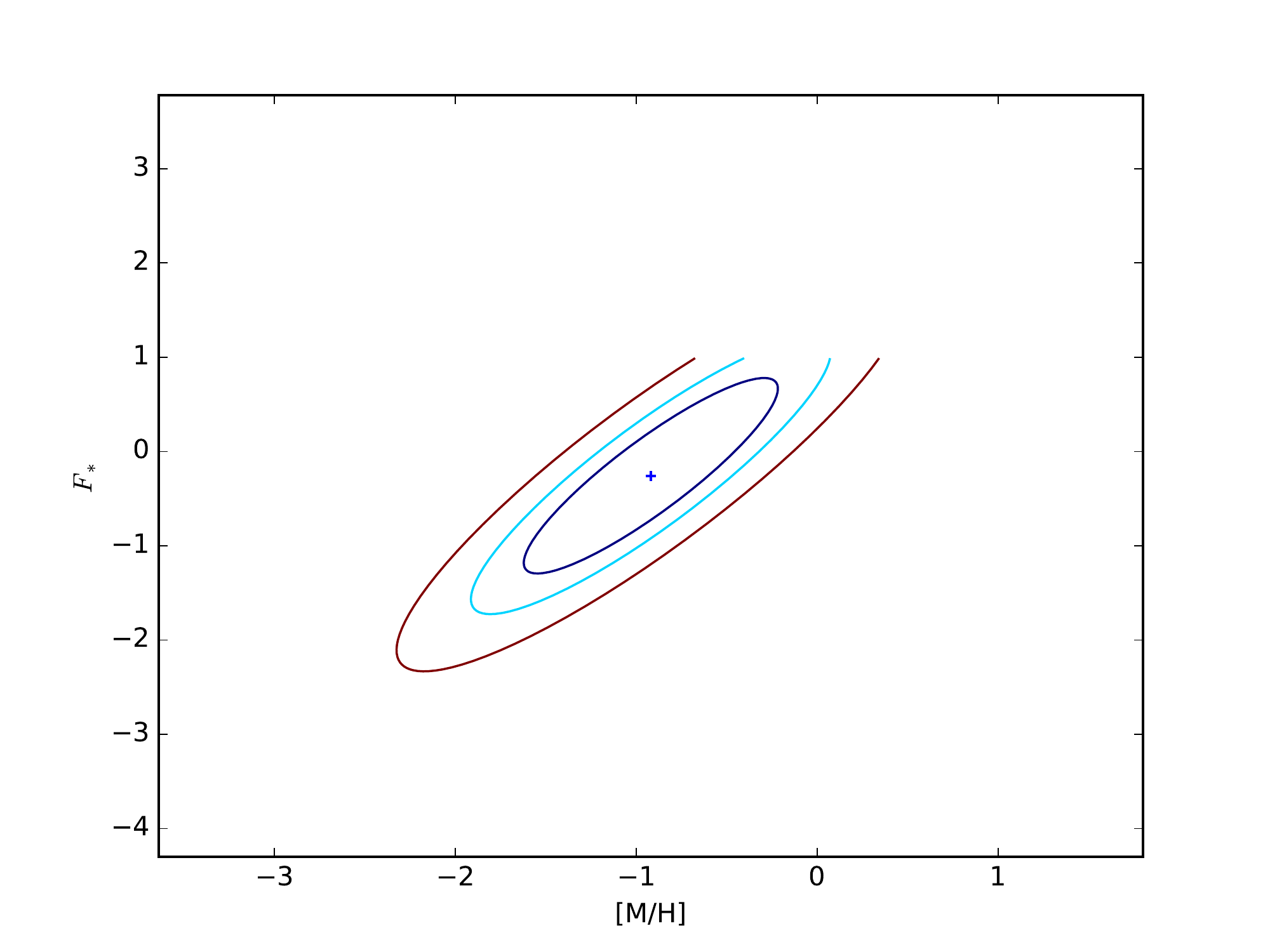}
\caption{Confidence regions for $F_*$ and [M/H] for GRB 050401.}
\label{contours050401}
\end{figure}
\begin{figure}[h]
\centering
\includegraphics[width=9cm]{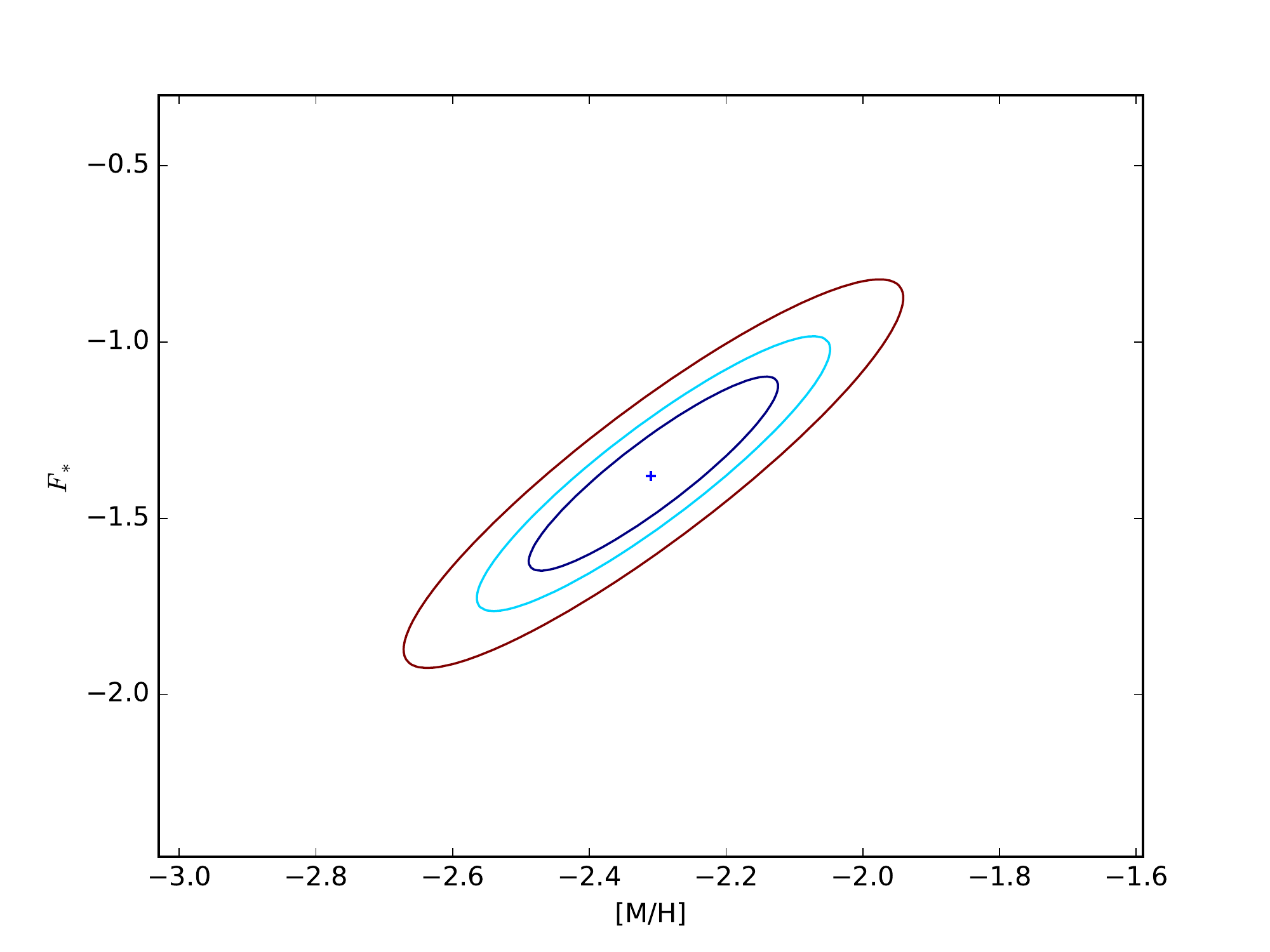}
\caption{Confidence regions for $F_*$ and [M/H] for GRB 050730.}
\label{contours050730}
\end{figure}
\begin{figure}[h]
\centering
\includegraphics[width=9cm]{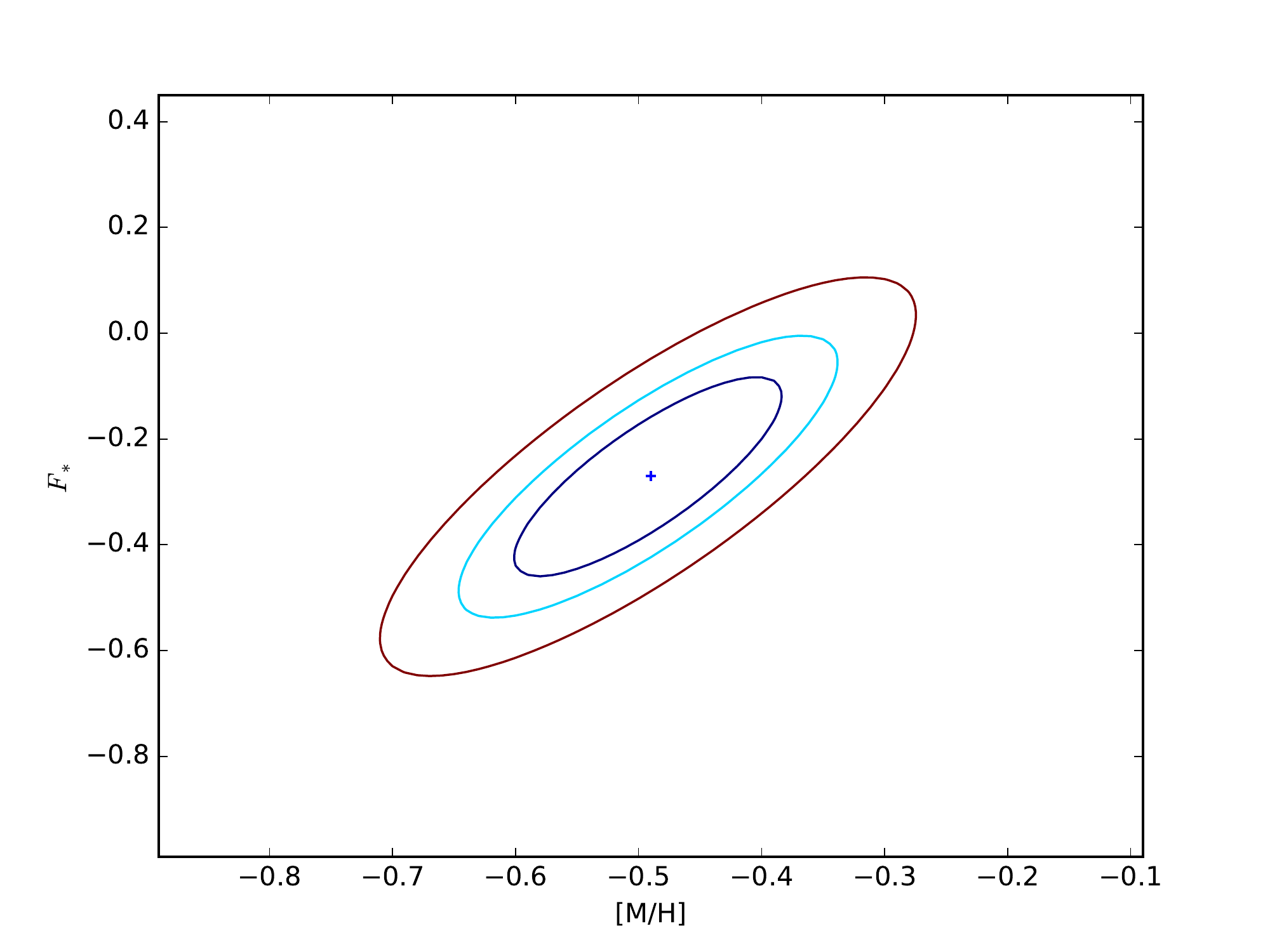}
\caption{Confidence regions for $F_*$ and [M/H] for GRB 050820A.}
\label{contours050820}
\end{figure}
\begin{figure}[h]
\centering
\includegraphics[width=9cm]{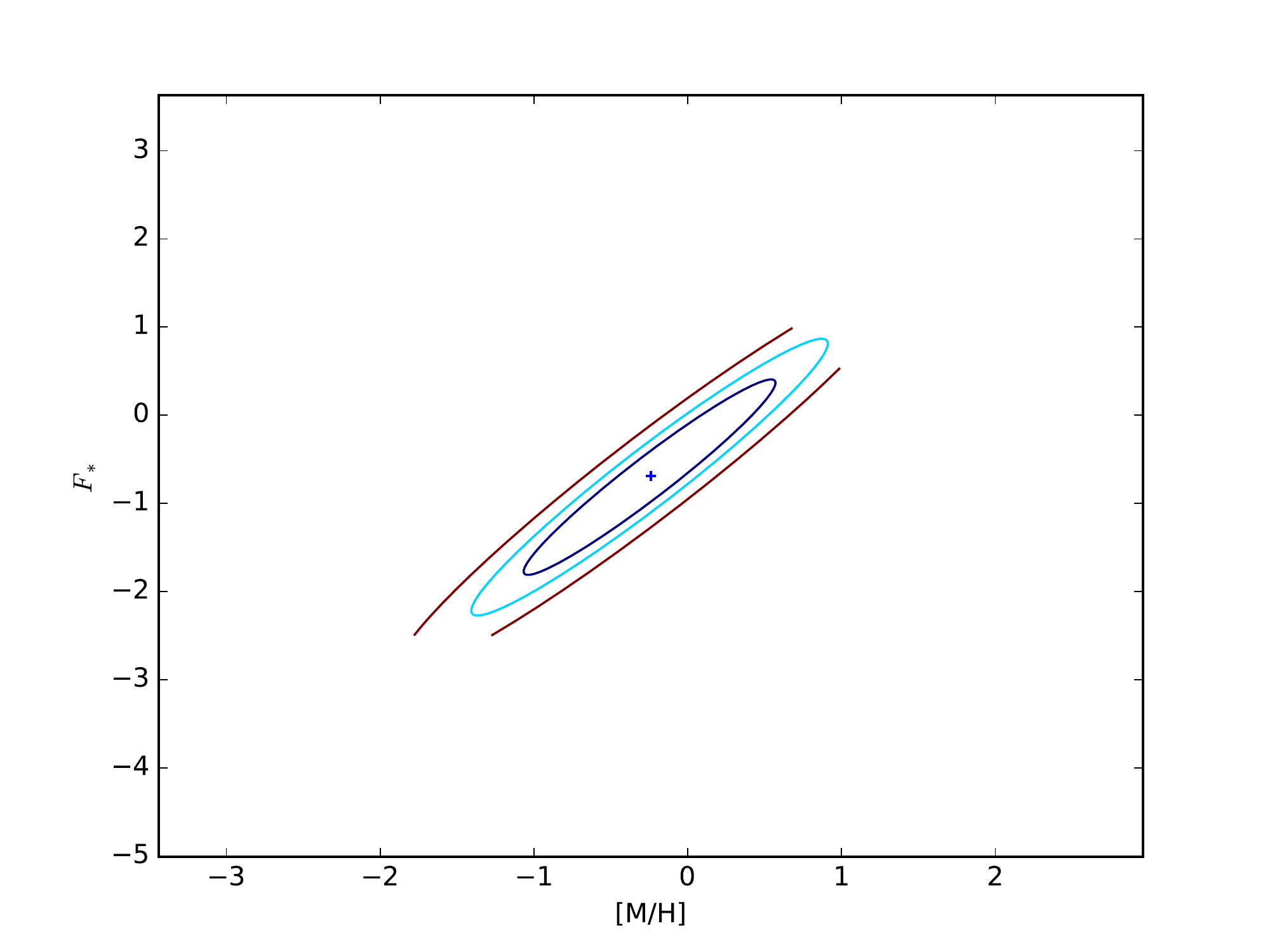}
\caption{Confidence regions for $F_*$ and [M/H] for GRB 070802.}
\label{contours070802}
\end{figure}
\begin{figure}[h]
\centering
\includegraphics[width=9cm]{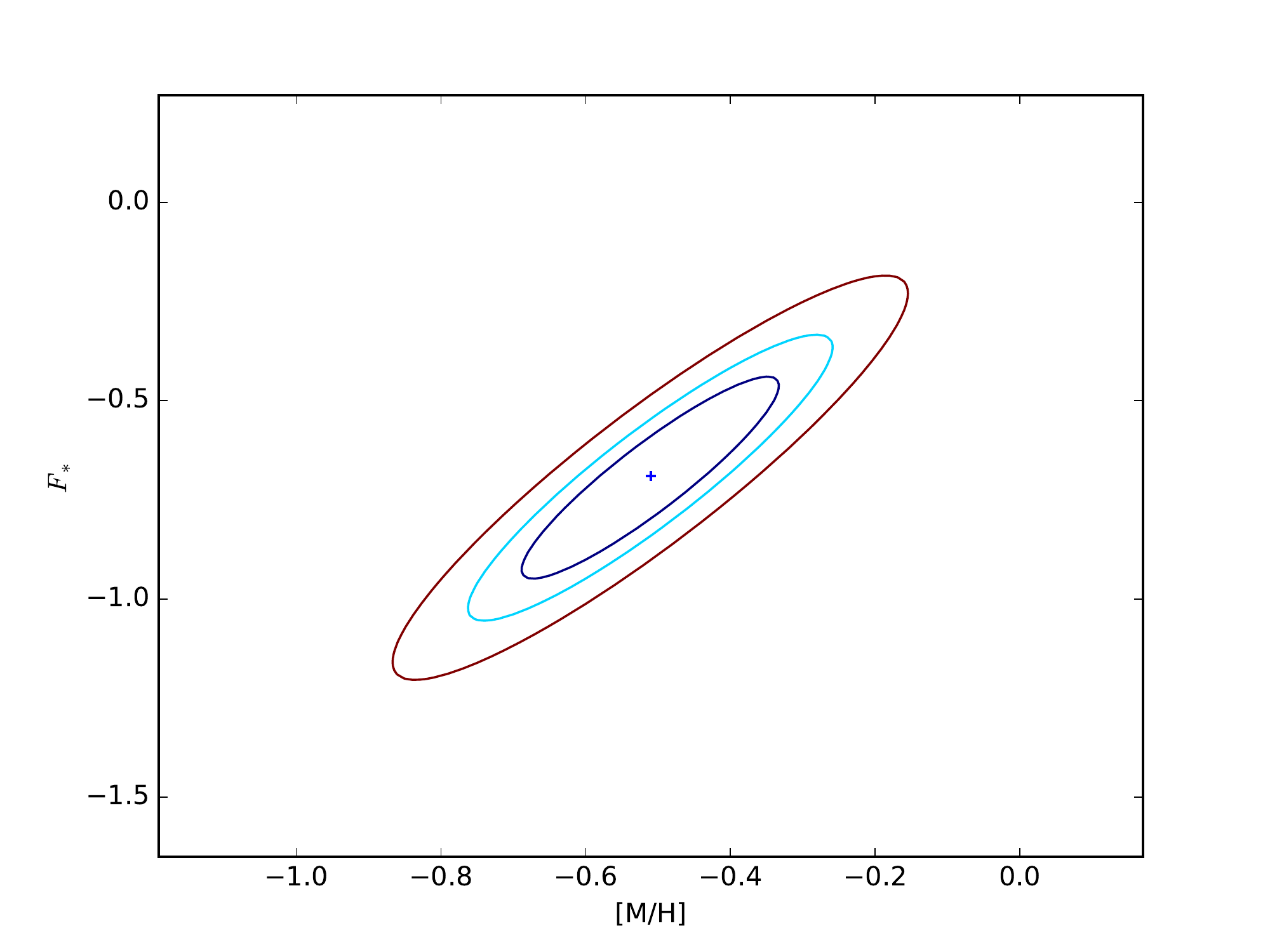}
\caption{Confidence regions for $F_*$ and [M/H] for GRB 081008.}
\label{contours081008}
\end{figure}
\begin{figure}[h]
\centering
\includegraphics[width=9cm]{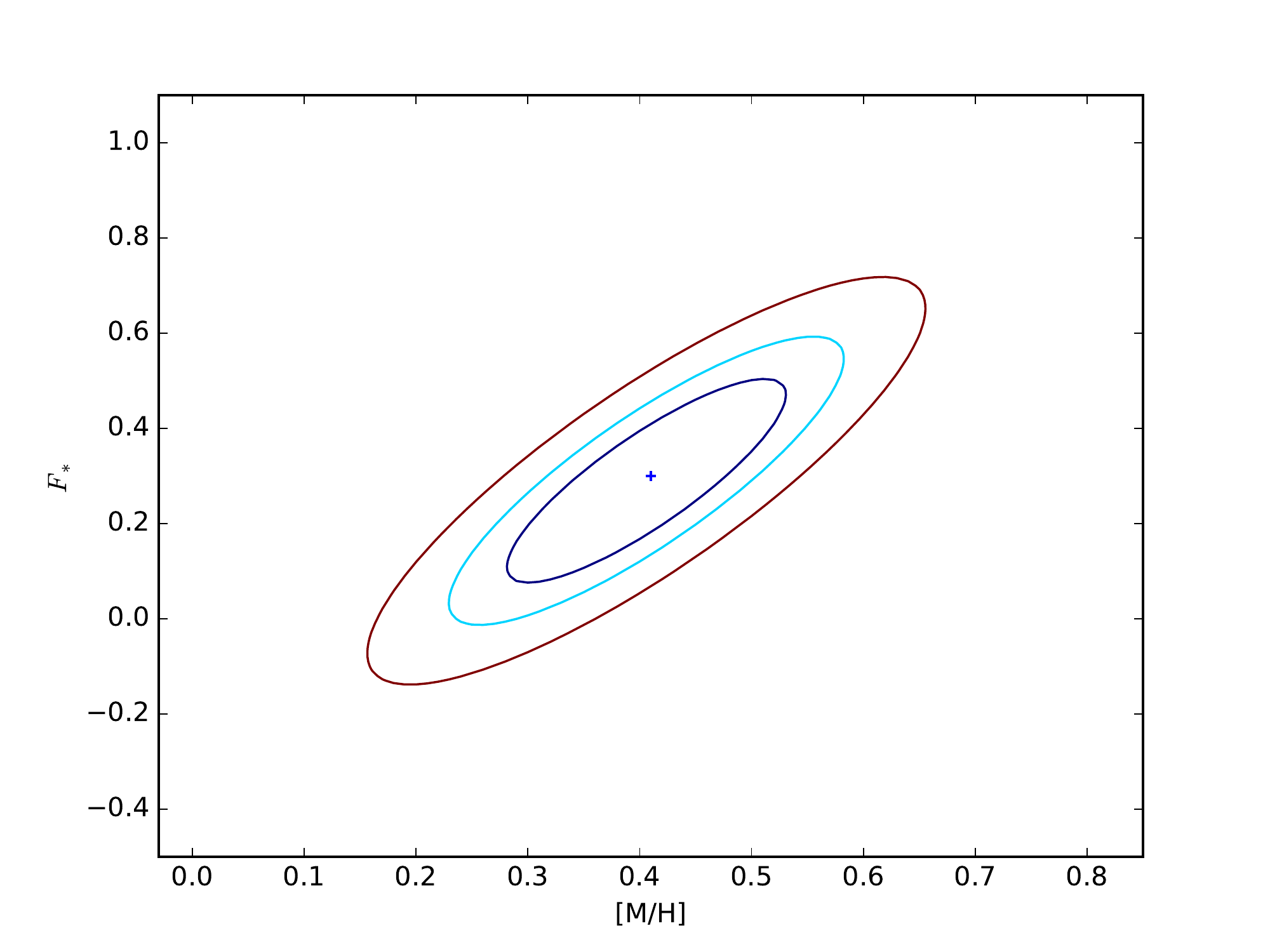}
\caption{Confidence regions for $F_*$ and [M/H] for GRB 090323.}
\label{contours090323}
\end{figure}
\begin{figure}[h]
\centering
\includegraphics[width=9cm]{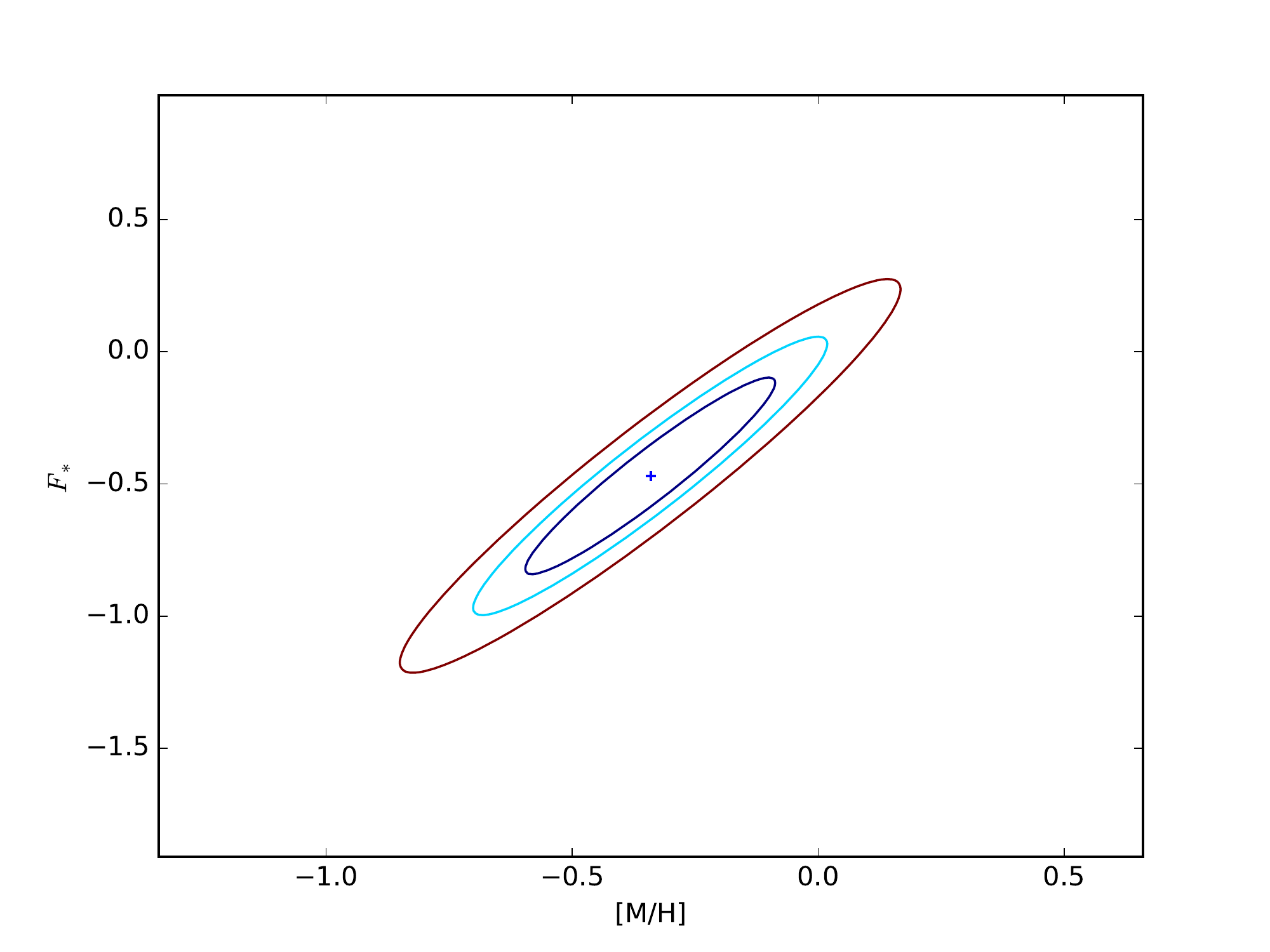}
\caption{Confidence regions for $F_*$ and [M/H] for GRB 090809F.}
\label{contours090809}
\end{figure}
\begin{figure}[h]
\centering
\includegraphics[width=9cm]{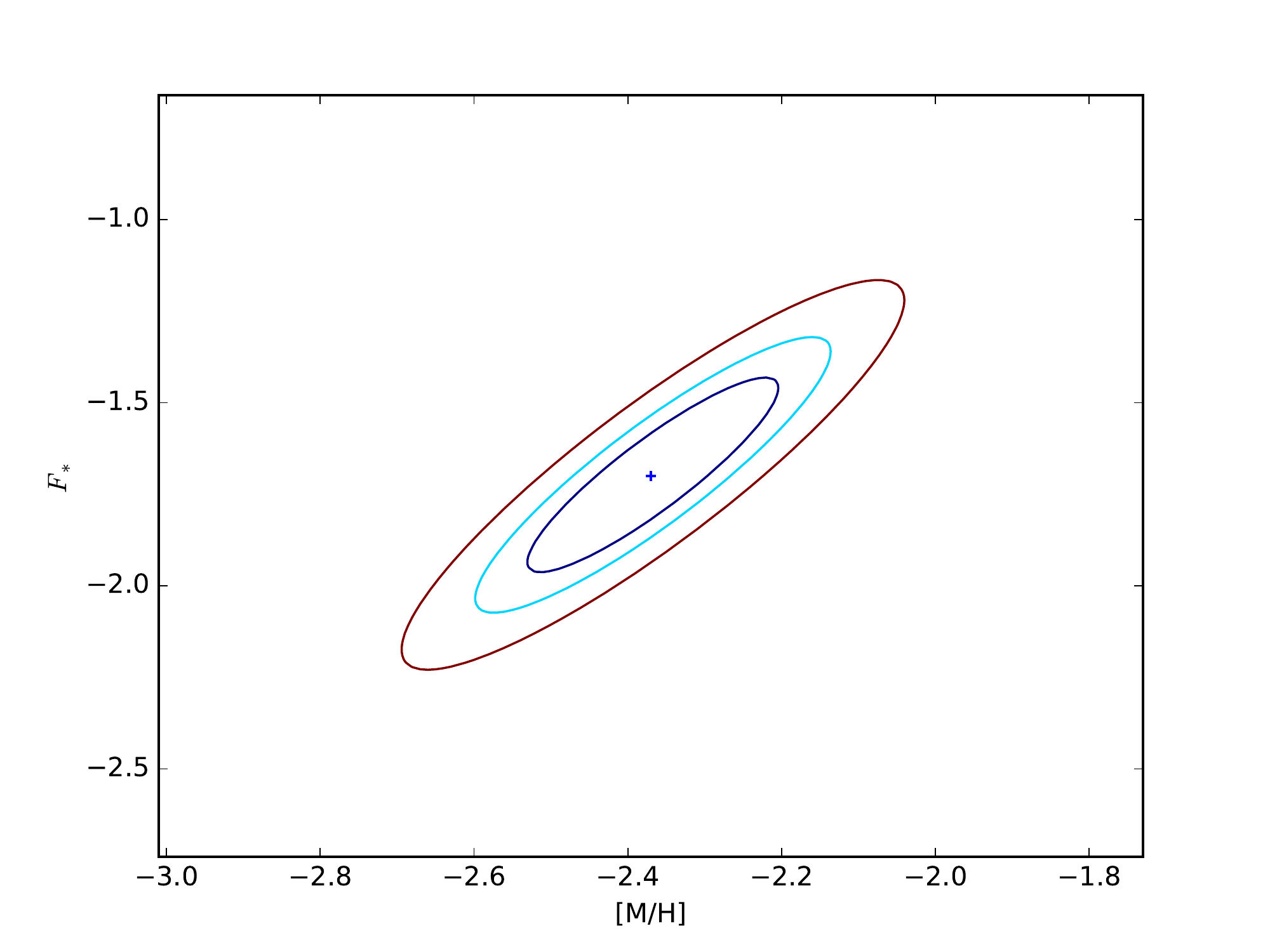}
\caption{Confidence regions for $F_*$ and [M/H]. for GRB 090926A}
\label{contours090926}
\end{figure}
\begin{figure}[h]
\centering
\includegraphics[width=9cm]{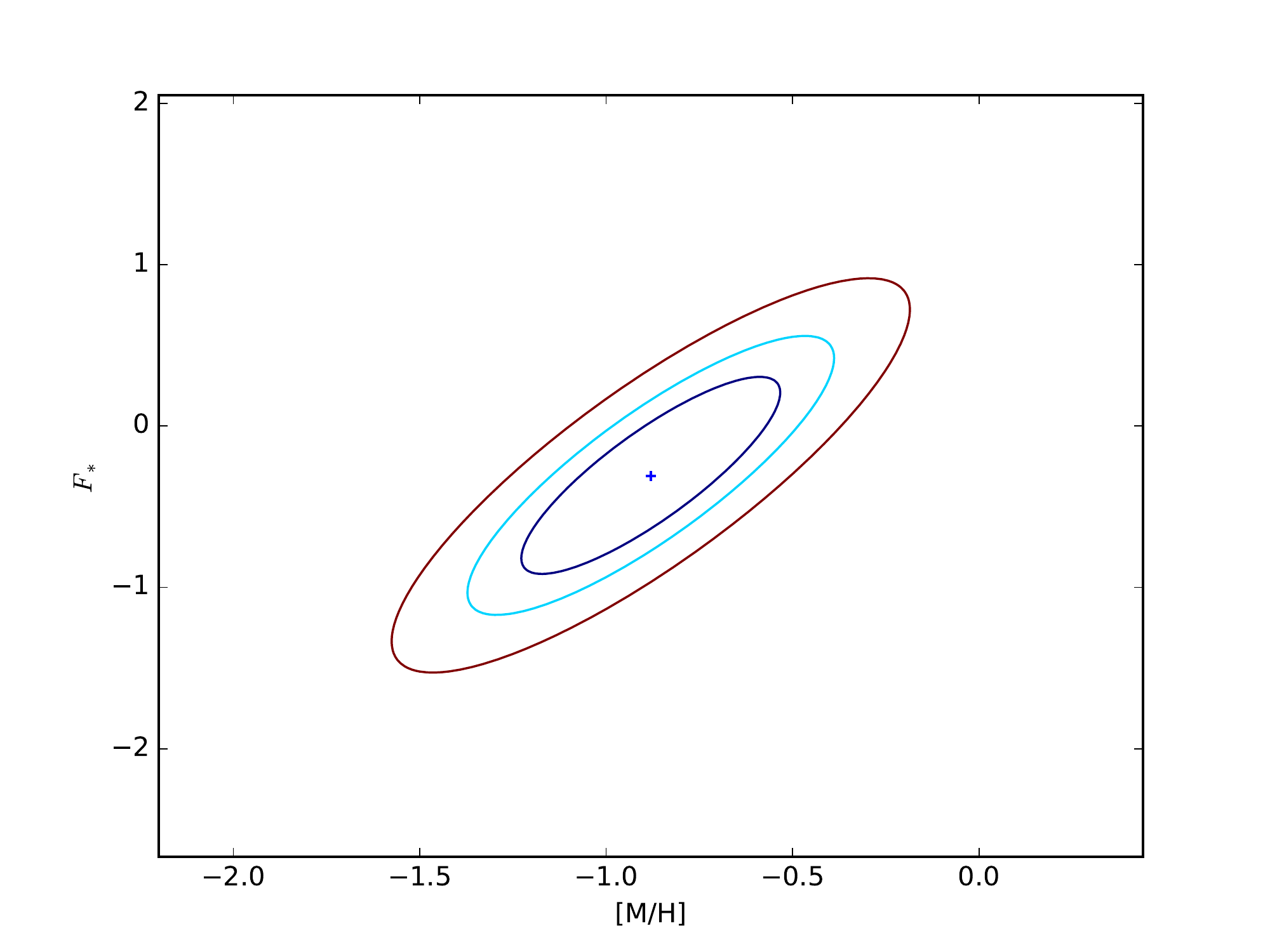}
\caption{Confidence regions for $F_*$ and [M/H] for GRB 100219A.}
\label{contours100219}
\end{figure}
\begin{figure}[h]
\centering
\includegraphics[width=9cm]{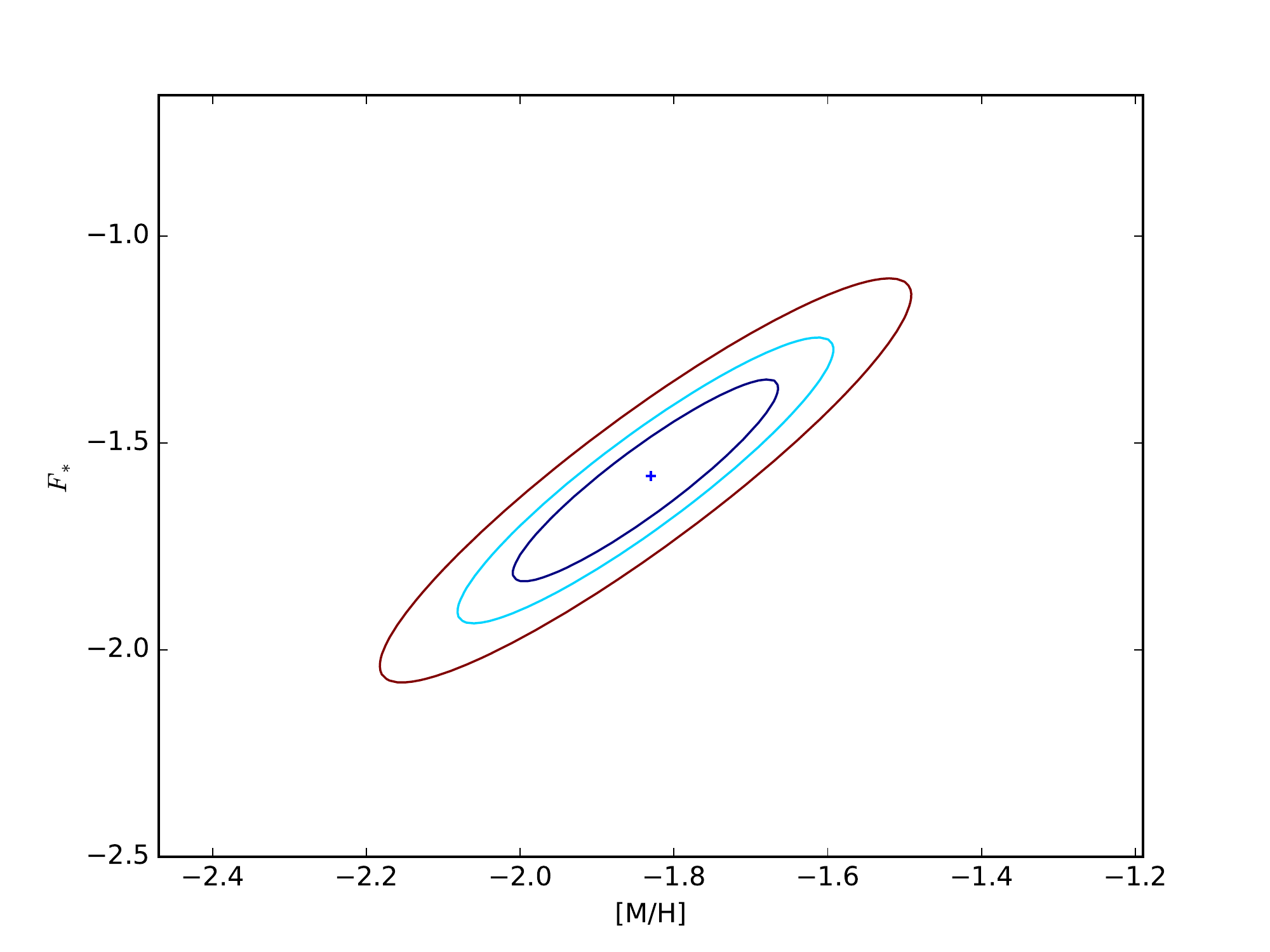}
\caption{Confidence regions for $F_*$ and [M/H] for GRB 111008A.}
\label{contours111008}
\end{figure}
\begin{figure}[h]
\centering
\includegraphics[width=9cm]{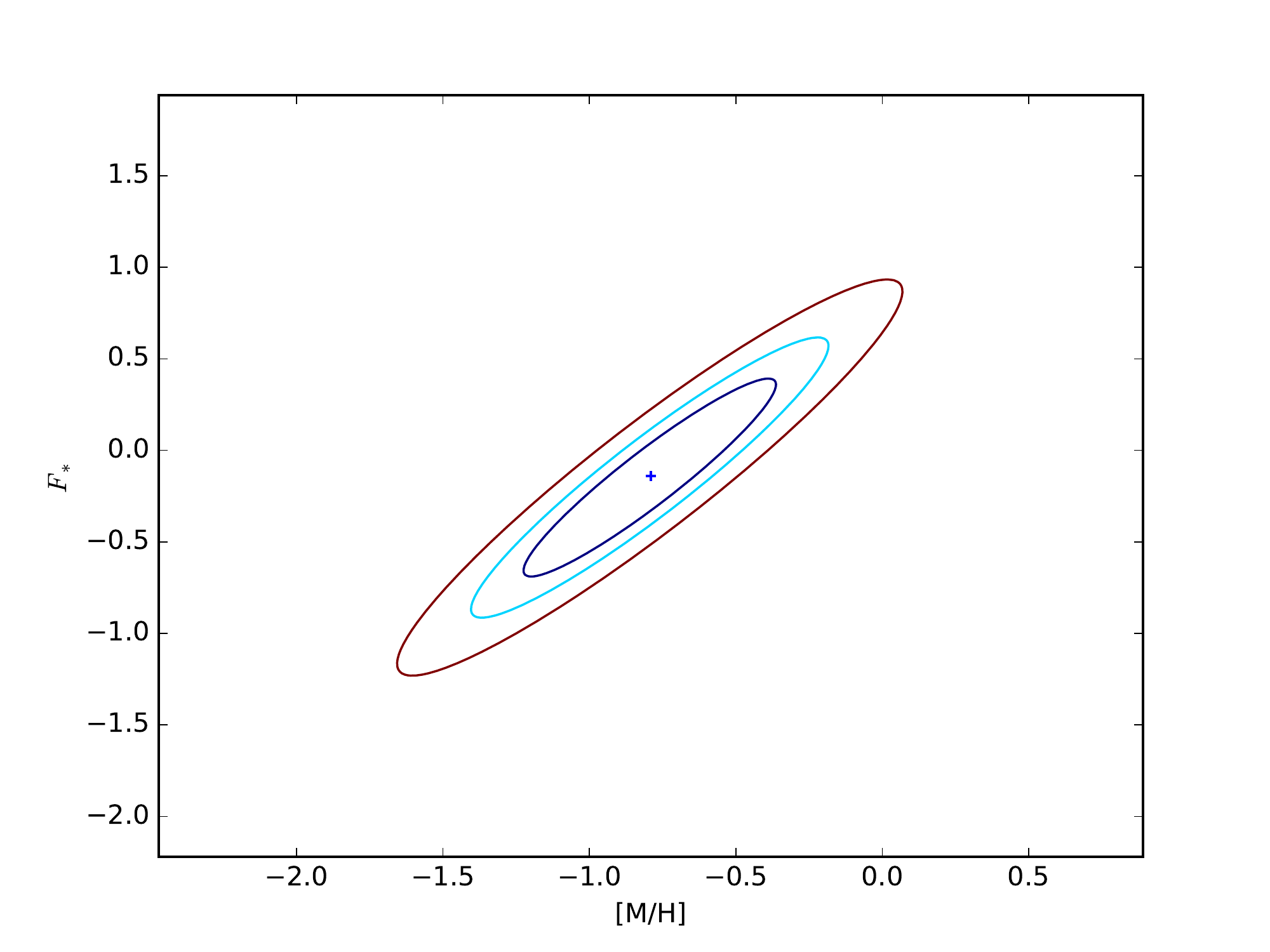}
\caption{Confidence regions for $F_*$ and [M/H] for GRB 120119A.}
\label{contours120119}
\end{figure}
\begin{figure}[h]
\centering
\includegraphics[width=9cm]{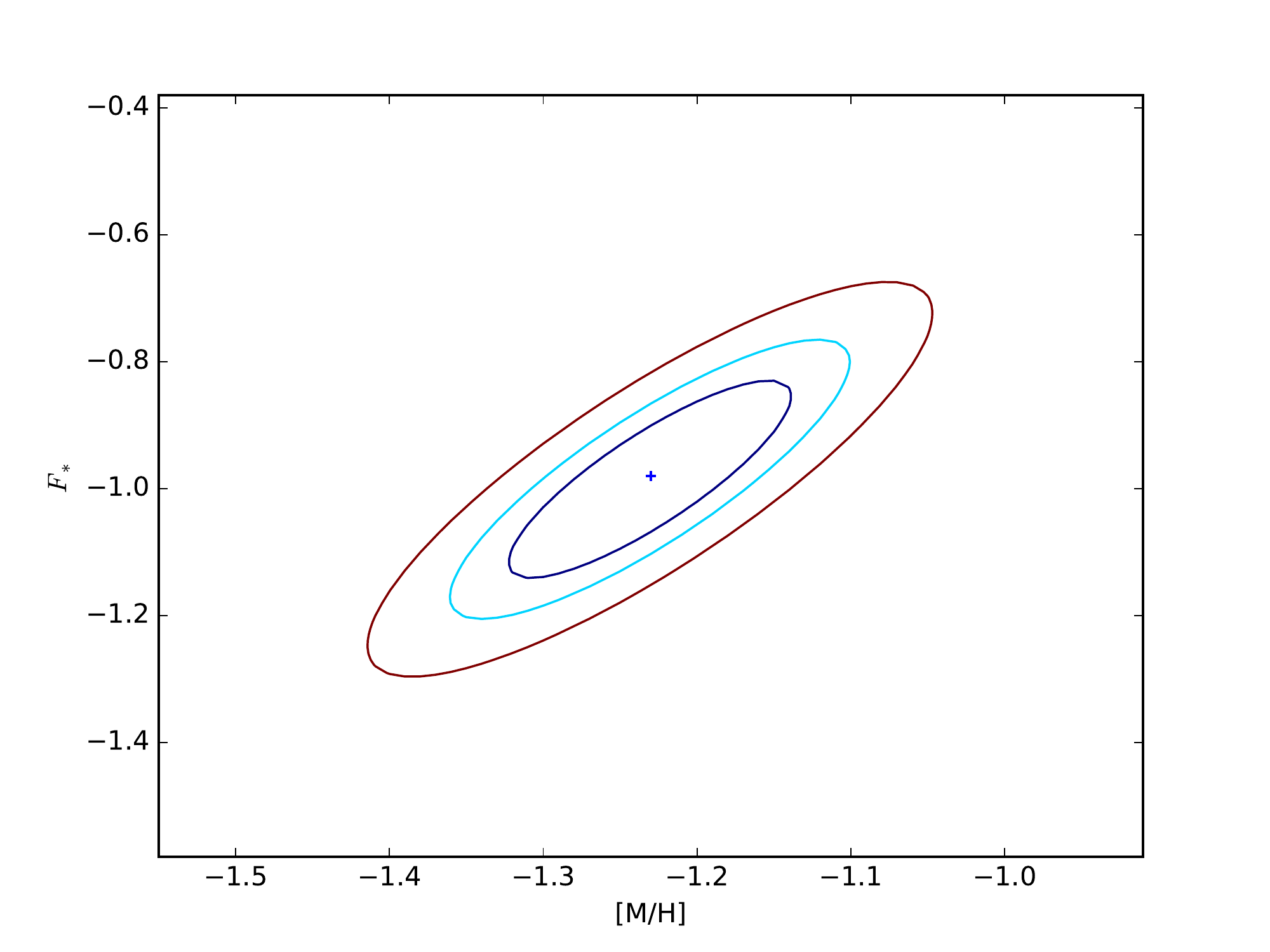}
\caption{Confidence regions for $F_*$ and [M/H] for GRB 120327A.}
\label{contours120327}
\end{figure}
\begin{figure}[h]
\centering
\includegraphics[width=9cm]{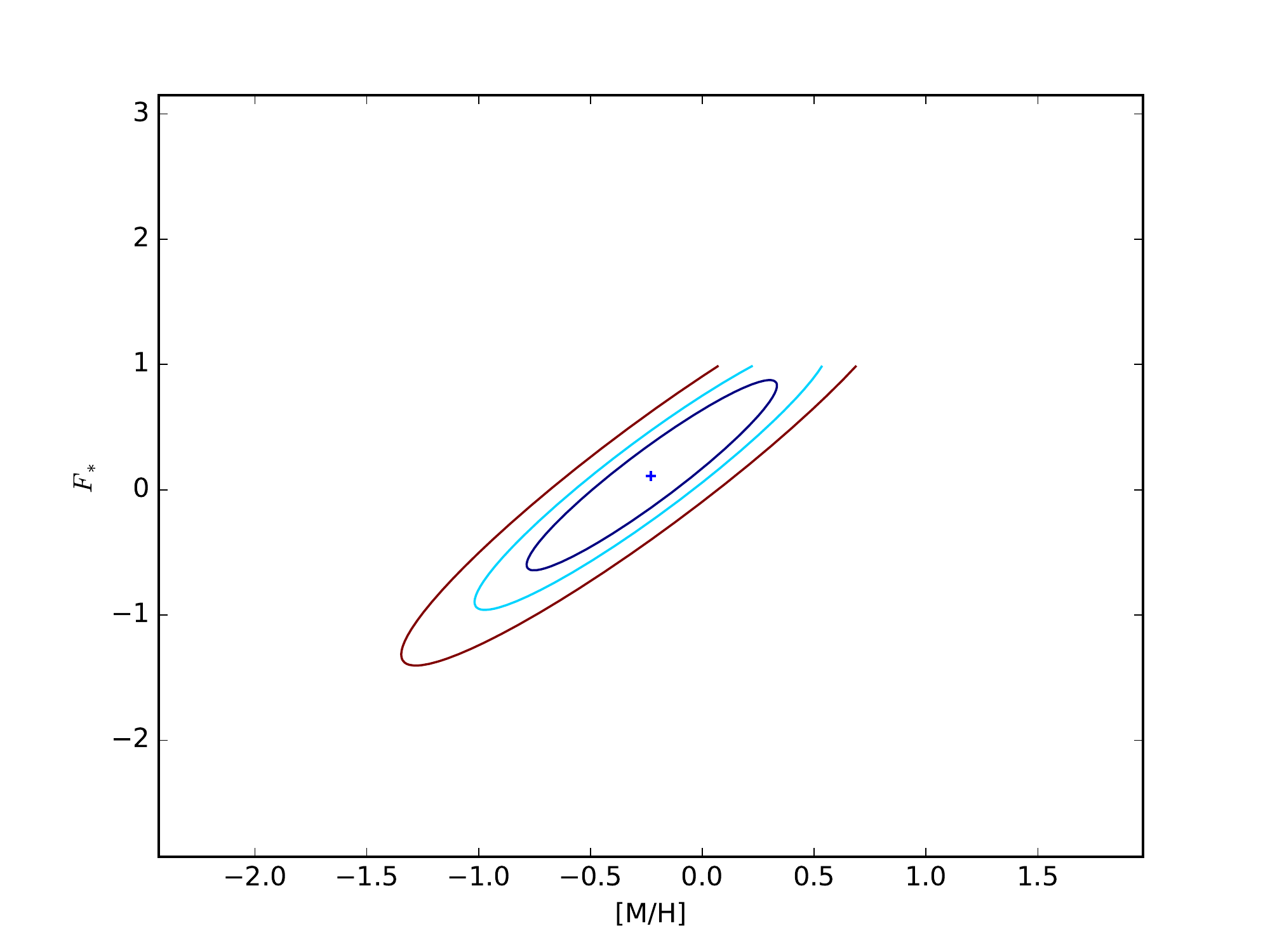}
\caption{Confidence regions for $F_*$ and [M/H] for GRB 120716A.}
\label{contours120716}
\end{figure}
\clearpage
\begin{figure}[h]
\centering
\includegraphics[width=9cm]{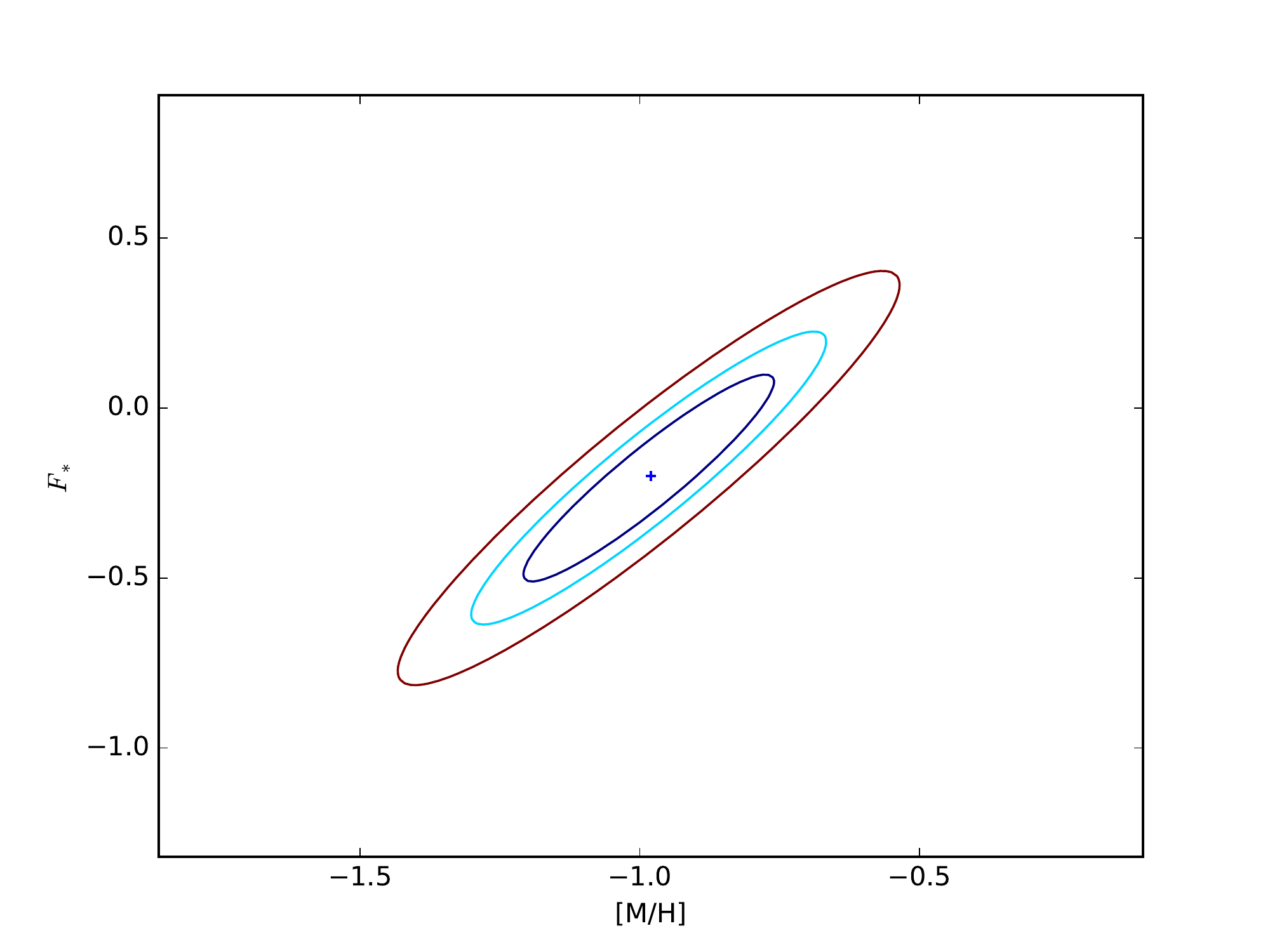}
\caption{Confidence regions for $F_*$ and [M/H] for GRB 120815A.}
\label{contours120815}
\end{figure}
\begin{figure}[h]
\centering
\includegraphics[width=9cm]{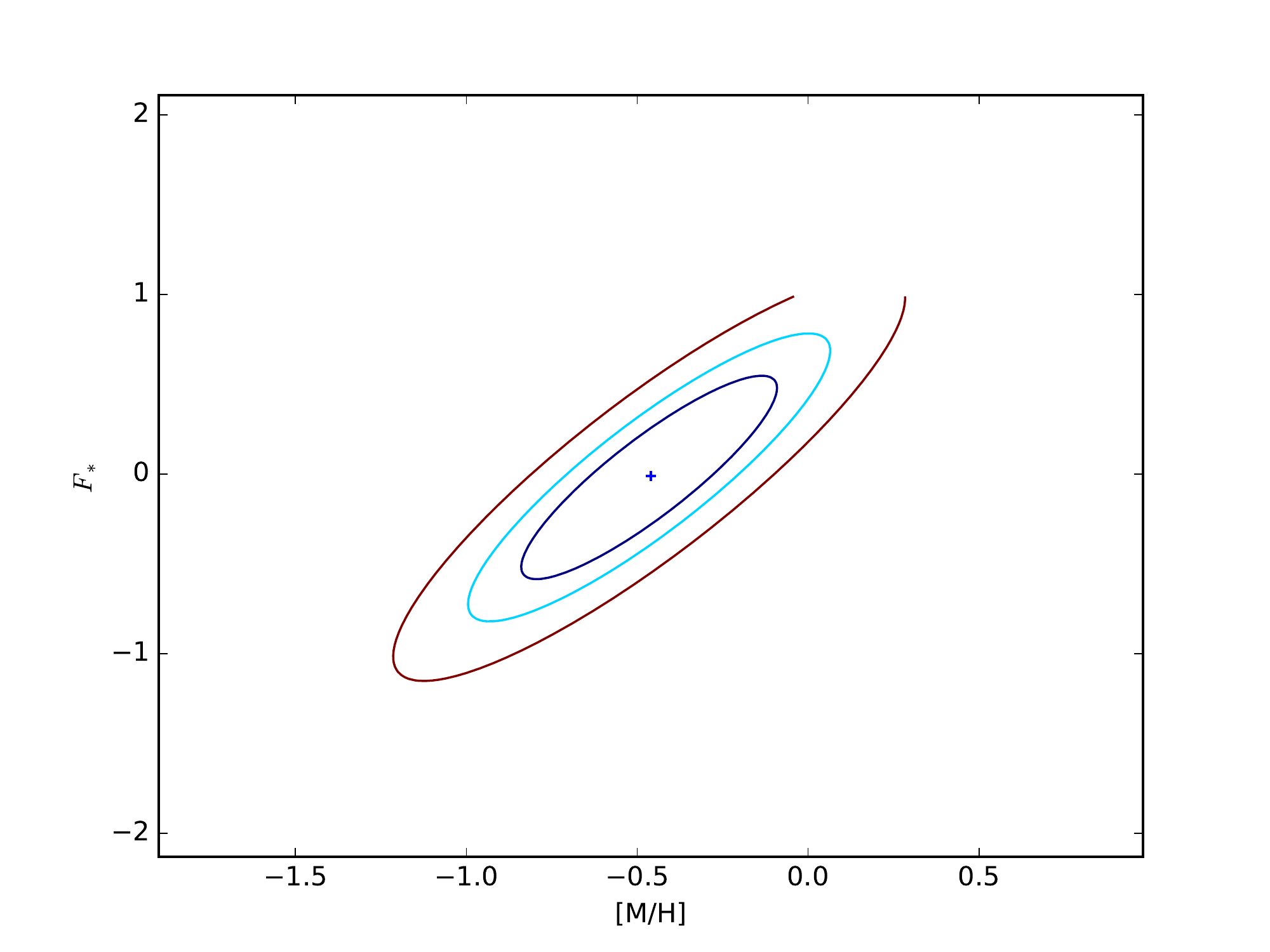}
\caption{Confidence regions for $F_*$ and [M/H] for GRB 120909A.}
\label{contours120909}
\end{figure}
\begin{figure}[h]
\centering
\includegraphics[width=9cm]{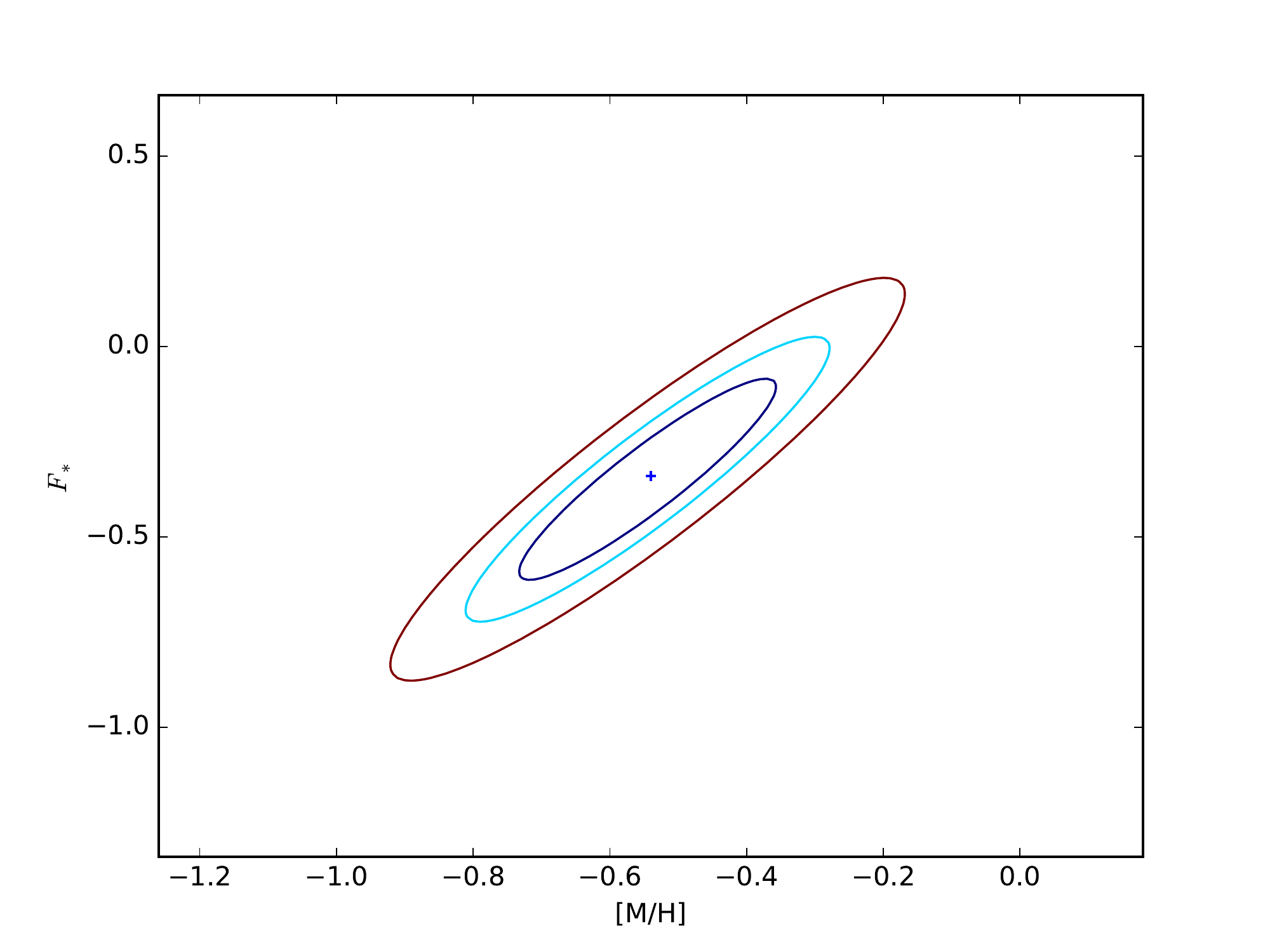}
\caption{Confidence regions for $F_*$ and [M/H] for GRB 121024A.}
\label{contours121024}
\end{figure}
\begin{figure}[h]
\centering
\includegraphics[width=9cm]{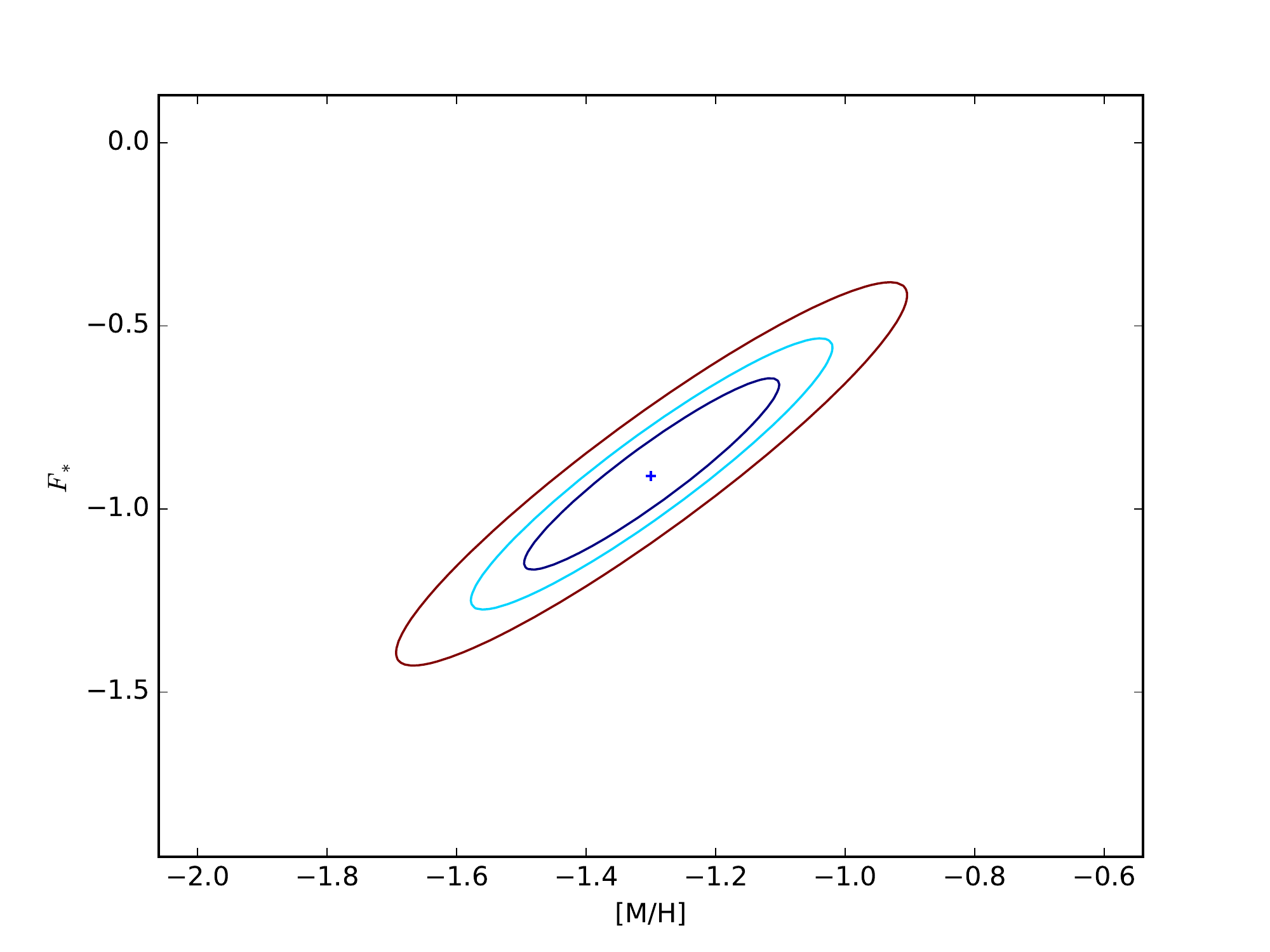}
\caption{Confidence regions for $F_*$ and [M/H] for GRB 130408A.}
\label{contours130408}
\end{figure}
\begin{figure}[h]
\centering
\includegraphics[width=9cm]{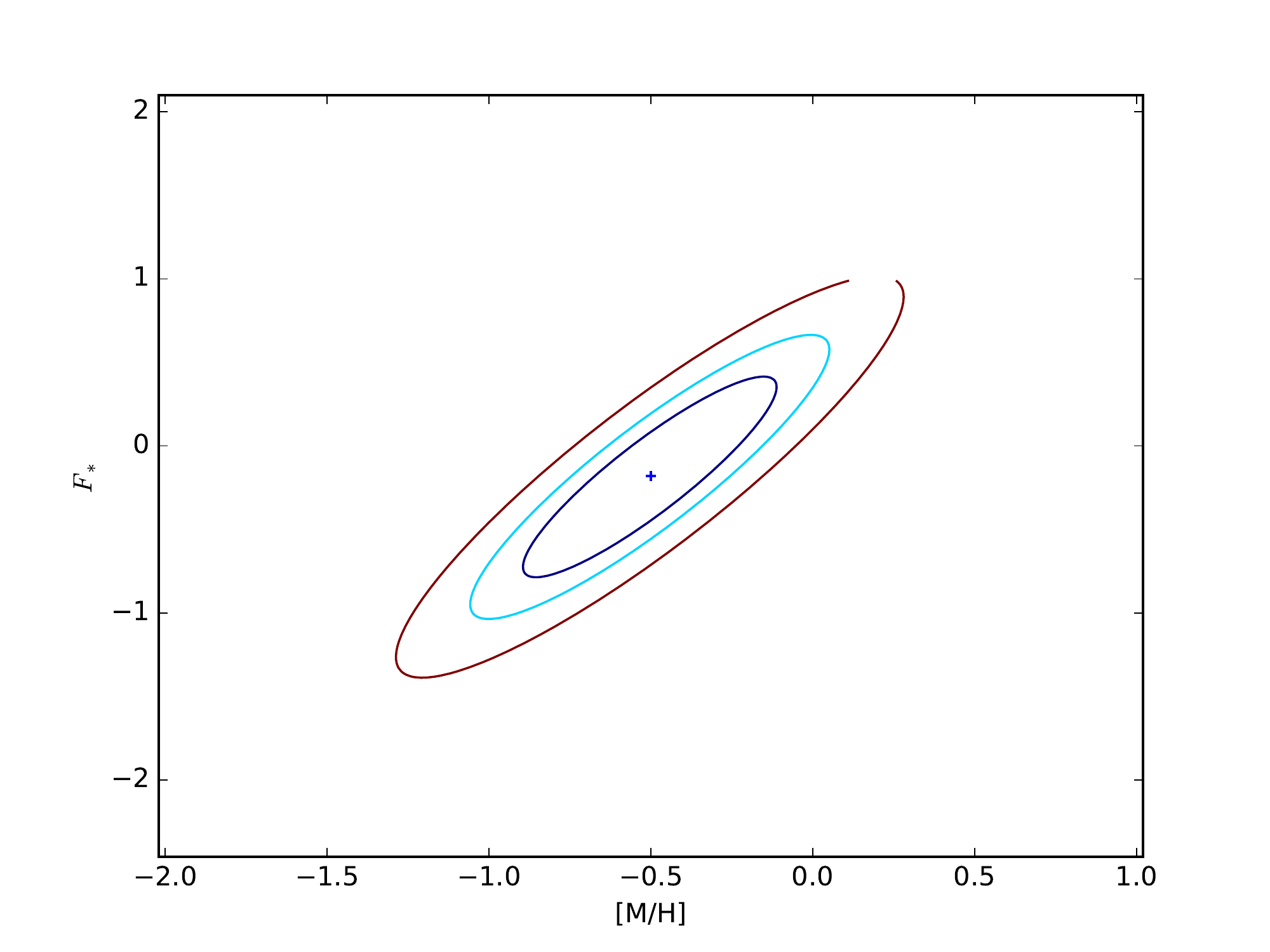}
\caption{Confidence regions for $F_*$ and [M/H] for GRB 141028A.}
\label{contours141028}
\end{figure}

\end{appendix}
\end{document}